\documentclass{article}
\pdfoutput=1

\usepackage{PRIMEarxiv}

\usepackage[utf8]{inputenc} %
\usepackage[T1]{fontenc}    %
\usepackage{hyperref}       %
\usepackage{url}            %
\usepackage{booktabs}       %
\usepackage{amsfonts}       %
\usepackage{nicefrac}       %
\usepackage{microtype}      %
\usepackage{lipsum}
\usepackage{fancyhdr}       %
\usepackage{graphicx}       %
\usepackage{ltablex}
\usepackage{enumitem}
\usepackage{float}
\usepackage{caption}
\graphicspath{{media/}}     %

\title{Assessing Generative AI value in a public sector context: evidence from a field experiment}

\author{
Trevor Fitzpatrick, Seamus Kelly, Patrick Carey\\
  Risk Analytics, Data and Supervisory Technology\\
  Supervisory Risk, Analytics and Data Directorate \\
  Central Bank of Ireland \\
  North Wall Quay, Dublin 1, Ireland\\
   \And
David Walsh, Ruairi Nugent\\
  Data Science and Analytics Team\\
  Supervisory Risk, Analytics and Data Directorate \\
  Central Bank of Ireland \\
  North Wall Quay, Dublin 1, Ireland\\
  }

\begin{document}
\maketitle

\begin{abstract}
The emergence of Generative AI (Gen AI) has motivated an interest in understanding how it could be used to enhance productivity across various tasks. We add to research results for the performance impact of Gen AI on complex knowledge-based tasks in a public sector setting. In a pre-registered experiment, after establishing a baseline level of performance, we find mixed evidence for two types of composite tasks related to document understanding and data analysis. For the Documents task, the treatment group using Gen AI had a 17\% improvement in answer quality scores (as judged by human evaluators) and a 34\% improvement in task completion time compared to a control group. For the Data task, we find the Gen AI treatment group experienced a 12\% reduction in quality scores and no significant difference in mean completion time compared to the control group. These results suggest that the benefits of Gen AI may be task and potentially respondent dependent. We also discuss field notes and lessons learned, as well as supplementary insights from a post-trial survey and feedback workshop with participants.
\end{abstract}


\section{Introduction}
The emergence of Generative AI (Gen AI) has motivated an interest in understanding how it could be used to enhance productivity across tasks in various professions. The potential productivity or efficiency improvements at a macroeconomic level \cite{Acemoglu2024} and at a microeconomic level within organisations is an emerging research focus.  Gen AI may have a more profound effect on how people work because of a potential range of more general capabilities that makes them multi-use as opposed to  traditional machine learning models which tend to be single use. However, this is more of a future possibility and not a present reality across most sectors and roles. There is much uncertainty as to the overall impact of Gen AI adoption. First, there are substantial questions of the ability of Gen AI to successfully execute complex tasks to a level of human performance. Second, there is a question as to whether the costs of Gen AI will decline sufficiently to make its adoption affordable across organisations, compared to existing methods of working \cite{Nathan:etal:2024}. Third, the nature of some types of public sector knowledge work means that existing studies may or may not be applicable to a public sector context. 

One way to navigate uncertainty is through gathering evidence via experimentation. \cite{Brynjolfsson:etal:2023} conducted one of the first studies that sought to understand how AI tools may change how workers perform and the impacts on their jobs. In a customer support setting, access to an AI-based conversational tool increased productivity by 14\% on average compared to the control of no AI access.  Research by \cite{NoyZhang2023} focused on incentivised writing tasks using university educated professionals. These tasks included writing press releases, short reports, analysis plans, and emails. These were designed to resemble real tasks performed in these occupations. The results suggest using ChatGPT compared to a control of no Gen AI assistance substantially raised productivity. The average time taken decreased by 40\% and output quality rose by 18\%. 

Using a variation on the design used in \cite{NoyZhang2023}, a very influential study of knowledge workers (management consultants) by \cite{Dellacqua:etal:2023} found that workers using AI completed 12\% more tasks, completed tasks 25\% more quickly, with more than 40\% higher quality compared to the control group, when performing tasks to create new product ideas focused on creativity, analytical thinking, and persuasive writing. The study also found that using AI significantly reduced performance compared to the control on a business problem-solving task involving analysing quantitative data, customer interviews, and persuasive writing. The authors suggest that a better understanding of user archetypes will support effective AI adoption.

However, these potential productivity or efficiency effects depend crucially on the context of the organisations in which people work. In this study, we assess the value of Gen AI in knowledge driven public sector organisations such as a central bank/financial regulation authority. The existing literature has begun to explore the potential promise of Gen AI in for-profit sectors, and given this is an emerging research area, there are gaps in understanding how this may impact public sector organisations.   

Unlike private enterprises, public sector organisations provide public services often through engaging in work where quality of decision making, its resilience, and effectiveness, as well as efficiency and productivity are balanced. These decisions may involve complex trade-offs, and there may be no one correct answer \cite{Bennett:etal:2023}. Public sector organisations generally operate under different accountability frameworks. By investigating the effects of Gen AI on tasks such as document comprehension and data analytics in a public sector setting, this study fills an important gap in understanding how Gen AI could be leveraged to improve effectiveness and enhance performance, including productivity in public policy driven environments.  

Central banks and financial regulators have access to and collect a rich variety of data and information to perform their functions. These functions include setting monetary policy, contributing to stable financial systems, and regulation of the financial sector. All of these mandates require knowledge work, specialised skills and capabilities, and Gen AI may help in facilitating more effective use of existing skills or widening access capabilities.    

However, there is still substantial uncertainty about the effects of Gen AI within organisations as this depends on the ability of organisations to adopt the technology, the pacing of this adoption, and the types of tasks to which Gen AI is applied. There are also questions around the economics of the provision of Gen AI solutions versus the cost of the status quo. In addition, as highlighted by \cite{Haslberger:etal:2023}, there could be skill disparities among workers, affecting the impact of adoption and use of AI. 

To assess the potential value of Gen AI to the Central Bank of Ireland (CBI), we conducted a small scale \href{https://www.socialscienceregistry.org/trials/13347}{pre-registered randomised control trial}. This involved two representative tasks – a document comprehension and data analytics task among 143 staff, representing about 7\% of the staff population of the CBI.

These tasks are typical of various types of knowledge work roles in central banking - obtaining information from complex documents, contextualising information and leveraging tacit knowledge, and providing answers to questions is common across many roles. Data analysis is important for economic, financial system, and policy related roles. Gen AI offers the potential for improving performance on both types of tasks for a wide variety of roles.  

Our results suggest Gen AI can significantly boost performance and productivity, but it depends on the task. For the document comprehension task, use of Gen AI resulted in a 17\% improvement in the first outcome measure – a quality score –and a 34\% improvement in the second measure – mean completion time – compared to the control group. For the Data task, the treatment group using Gen AI experienced a 12\% reduction in quality scores and no significant difference in mean completion time compared to the control group. 

In our results discussion, we examine some reasons why we may be observing this result. We also illustrate the variation in performance across questions within each task. Less complex questions resulted in greater performance difference between treatment and control, and more complex questions resulted in the reverse. These and other findings suggest further rich research avenues around the benefits of Gen AI in a public sector context. Finally, we share some field notes and experiences from this trial, including how it can be improved upon in future.

\section{Methods}
\label{sec:methods}
\begin{figure}[htbp]

\centering
\includegraphics[width=\textwidth]{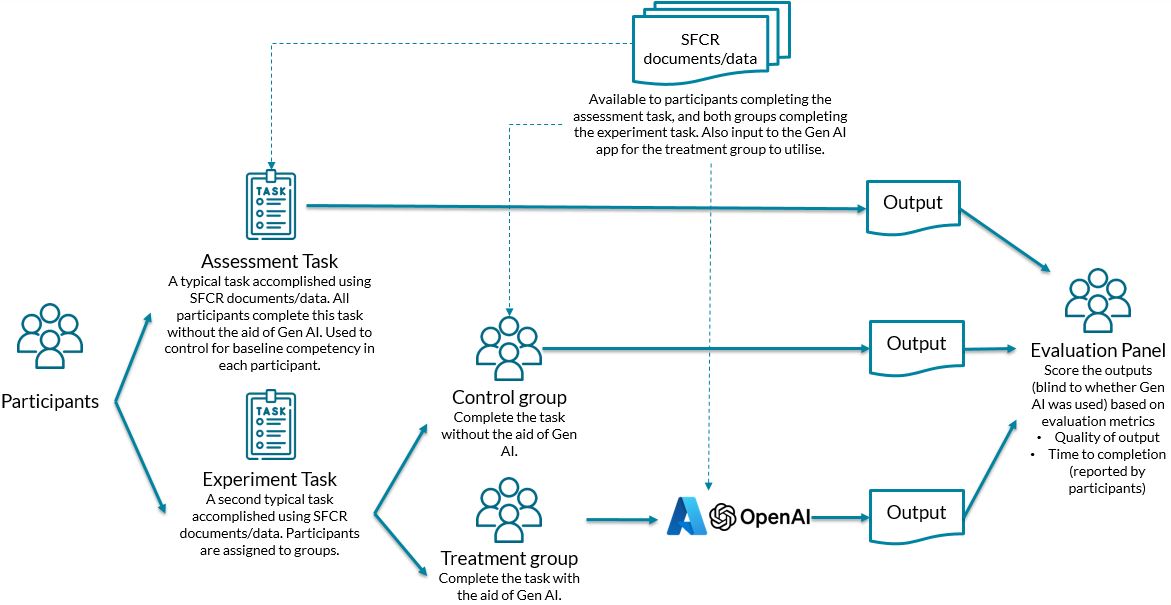}
\caption{Experiment Design}
\label{fig:experiment_design}
\end{figure}

Figure \ref{fig:experiment_design} illustrates the design of the experiment, which is a simplified version of that used by \cite{NoyZhang2023} and \cite{Dellacqua:etal:2023}.

Each potential participant was asked to complete an enrolment survey and provide some demographic, educational, and other control information described further in Appendix \ref{sec:enrolmentsurvey}. Approximately 208 staff registered interest in the trial. Participants could register their preference to participate in either the Data or Documents task. Every participant was assigned to their preferred task. Participants then completed an assessment task without the aid of a Gen AI tool to measure their baseline competency in performing either a Data or Documents task depending on their task assignment. After completion of the assessment task, participants were then randomly assigned to either the treatment group receiving access to a Gen AI tool to undertake the experiment task, or the control group that did not. On completion of the tasks, the Documents task submissions were scored for the primary outcome quality measure by two human evaluators based on a pre-defined rubric, and the average score across the two evaluators taken. The Data task submissions were scored by human graders against a known ground truth.

\subsection{Tasks}

Each task consisted of a series of questions that could be answered through review of insurance companies’ Solvency and Financial Condition Report (SFCR) documents for the Documents task, or analysis of a large insurance dataset for the Data task (see Appendix \ref{sec:appendixtasks} for more details on the tasks). The control group for the Documents task received the document in a PDF file. The control group for the Data task received an Excel file with data and pivot tables. For the treatment groups, the Gen AI tools were web applications utilising GPT-4o configured to provide answers for the specific task (see Appendix \ref{sec:apps} for more details on the applications). For the Data task, questions were provided and answers were submitted in a document. For the Documents task, questions were provided and answers were submitted in a survey tool. No background in insurance was assumed, and participants were provided terminology and definition sheets.

\subsection{Outcome measures}

The primary outcome variable is the quality of responses to the tasks. For the Documents task, quality was assessed by an independent panel of human graders using an evaluation rubric, as detailed in Appendix \ref{sec:documentstasks}. The response to each question was evaluated by two graders, and the mean of the two grades was taken to provide a quality score for each question. The quality scores for all questions were then averaged to derive an overall quality score for the task. For the Data task, quality was assessed by comparing the responses to known ground truth values, as detailed in Appendix \ref{sec:datatasks}. For both tasks, evaluation was performed blind to whether the participant was in the treatment or control group. 

The secondary outcome variable is the time taken to complete the task. For both the Documents and Data tasks this was self-reported - participants were asked to record the time taken to complete the task, and to include this with their response submissions.

\section{Results}
\label{sec:mainresults}
Overall, the results paint a mixed picture. For the Documents task, the treatment group using Gen AI had a 17\% improvement in quality scores and a 34\% improvement in task completion time. For the Data task, the Gen AI treatment group experienced a 12\% reduction in quality scores and no significant difference in task completion time compared to the control group. 

Putting these results into context compared to other studies with similar designs but differing tasks, the observed quality improvement for the Documents task is slightly lower than the quality improvement in \cite{NoyZhang2023} and significantly lower than reported in \cite{Dellacqua:etal:2023}. For the Documents task, improvement in time to completion is within the range reporting in \cite{NoyZhang2023} and \cite{Dellacqua:etal:2023}. For the Data task, there are limited results to which to compare our result in the literature. 

\begin{figure}[htbp]
\centering
\begin{minipage}{0.45\textwidth}
\centering
\includegraphics[width=\textwidth]{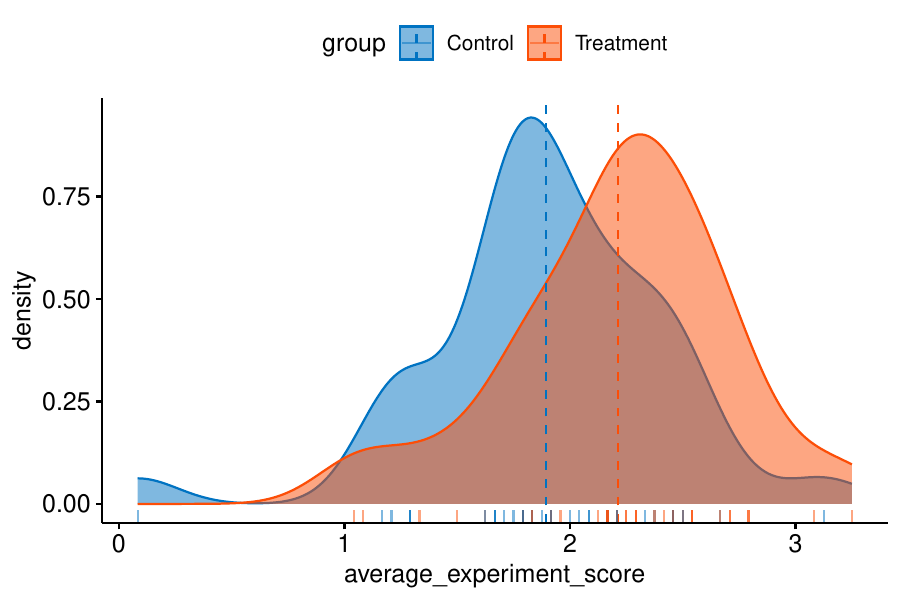}
\caption{Documents Task - Quality}
\label{fig:score_density}
\end{minipage}\hfill
\begin{minipage}{0.45\textwidth}
\centering
\includegraphics[width=\textwidth]{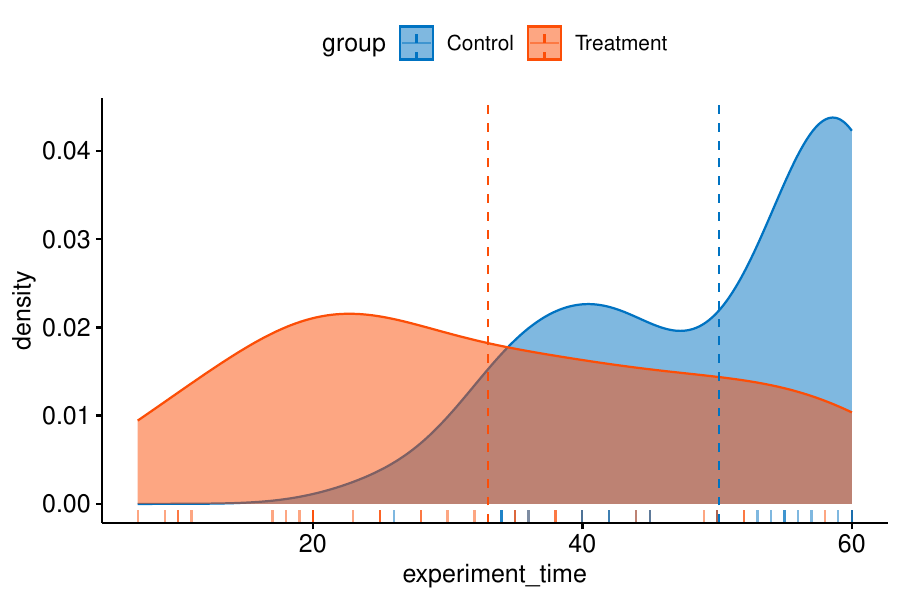}
\caption{Documents Task - Time}
\label{fig:score_time}
\end{minipage}
\caption*{\textbf{Note:} The figures show the full outcome variable distribution (Figure \ref{fig:score_density} - quality score, Figure \ref{fig:score_time} - time) in the Documents experiment task for treatment (red) and control (blue) groups. Dashed lines indicate the mean for each group.}
\end{figure}

\begin{figure}[htbp]
\centering
\begin{minipage}{0.45\textwidth}
\centering
\includegraphics[width=\textwidth]{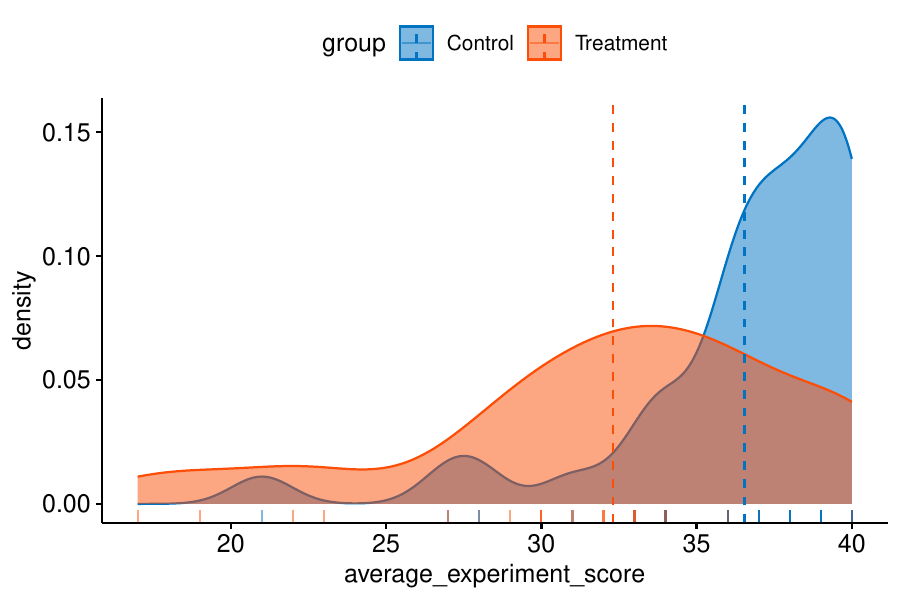}
\caption{Data Task - Quality}
\label{fig:density_quality}
\end{minipage}\hfill
\begin{minipage}{0.45\textwidth}
\centering
\includegraphics[width=\textwidth]{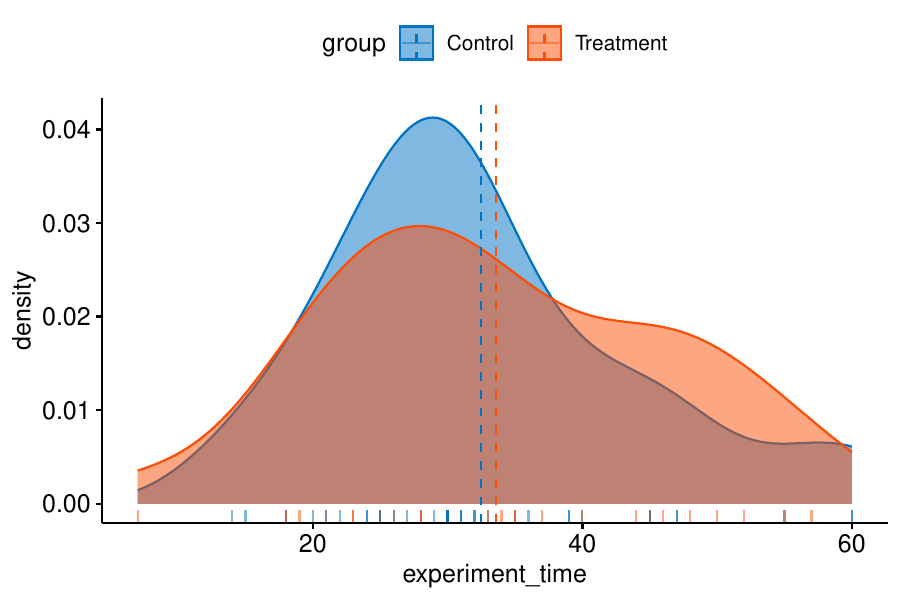}
\caption{Data Task - Time}
\label{fig:density_time}
\end{minipage}
\caption*{\textbf{Note:} The figures show the full outcome variable distribution (Figure \ref{fig:density_quality} - quality score, Figure \ref{fig:density_time} - time) in the Data experiment task for treatment (red) and control (blue) groups. Dashed lines indicate the mean for each group.}
\end{figure}

The following four subsections provide more detail on the results for each of the two tasks for the primary and secondary outcome. 

\subsection{Documents - Primary Outcome: Quality of Responses}

The primary outcome of interest was the quality of responses to the document comprehension task. Participants who had access to Gen AI demonstrated a small but significant improvement in response quality compared to the control group. The density plot (Figure \ref{fig:score_density}) shows the distribution of scores for both the treatment and control groups. The average quality score for the treatment group was 2.21 compared to 1.89 in the control group, a mean difference of 0.32. This represents a 17\% improvement in quality for the treatment group. A two-tailed t-test confirmed that this difference was significant (p < 0.05).

Table \ref{docsregression} presents the results of the analyses examining the impact of using Gen AI on the quality outcome variable across the three model specifications. Column 1 uses only the treatment variable, while Columns 2 and 3 progressively incorporate additional covariates, including enrolment data and the baseline assessment. 

In column 1, the Gen AI treatment leads to a statistically significant increase of 0.321 over the control, representing a 17\% improvement. In Column 2, after controlling for enrolment data covariates, the treatment effect remains positive but decreases slightly to 0.219, with marginal significance. In Column 3, which includes both enrolment data and the baseline assessment, the treatment effect is 0.227 and statistically significant (p < 0.05).

Across all specifications, the inclusion of additional covariates improves the model fit substantially, as reflected in the increasing adjusted-$R^2$ values (from 0.081 in Column 1 to 0.406 in Column 3). Model quality improves with the inclusion of these variables, with AIC decreasing from 116.4 in Column 1 to 102.3 in Column 3, indicating better model performance. While the treatment effect decreases slightly with the inclusion of enrolment and baseline assessment data, Gen AI consistently shows a positive effect on the outcome.

As a robustness check, we use two non-parametric estimators - Targeted Maximum Likelihood Estimators \cite{Vanderlann_tmle:2006,tmlepackage:2012} and Generalised Random Forests \cite{AtheyWagerGRF:2019} - which result in slightly larger effect size estimates compared to the linear model. These are shown with the associated standard errors in Table \ref{quality_effect_estimators_table}. The same set of covariates are included in both models as in Column 3 of Table \ref{docsregression}.

\begin{table}[h]  
    \centering  
    \caption{Documents Task - Comparison of linear treatment effect with machine-learning estimators}  
    \label{quality_effect_estimators_table}  
    \begin{tabular}{lcc}  
        \toprule  
        Estimator & Effect Estimate & Standard Error \\  
        \midrule  
        Linear    & 0.227 & 0.112 \\  
        TMLE      & 0.342 & 0.083 \\  
        GRF       & 0.300 & 0.112 \\  
        \bottomrule  
    \end{tabular}  
\end{table} 

\subsection{Documents - Secondary Outcome: Task Completion Time}
The secondary outcome of interest was the time taken to complete the document comprehension task. The density plot (Figure \ref{fig:score_time}) shows the distribution of time taken to complete the task for both the treatment and control groups. There was a substantial reduction in completion time for the treatment group compared to the control group. 

On average, participants in the treatment group completed the task in 33 minutes, which was significantly shorter than the control group's average completion time of 50 minutes. A mean difference of 17 minutes represents a 34\% improvement for the treatment group in average time to complete the task compared to the control group. The time difference was statistically significant (p < 0.05), suggesting the use of Gen AI reduced the time participants spent in answering the comprehension questions. Time to completion in our trial was self-reported, with a 60 minute time limit.

\subsection{Documents - Main Beneficiaries}

\begin{figure}[htbp]
\centering
\begin{minipage}{0.45\textwidth}
\centering
\includegraphics[width=\textwidth]{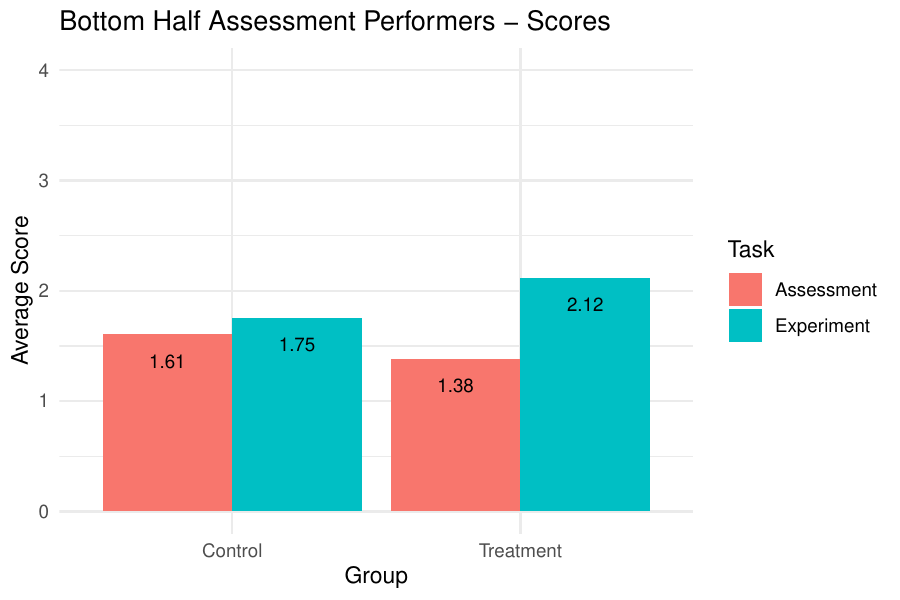}
\caption{Documents Task - Bottom-Half Assessment: Scores}
\label{fig:bh_baseline_score}
\end{minipage}\hfill
\begin{minipage}{0.45\textwidth}
\centering
\includegraphics[width=\textwidth]{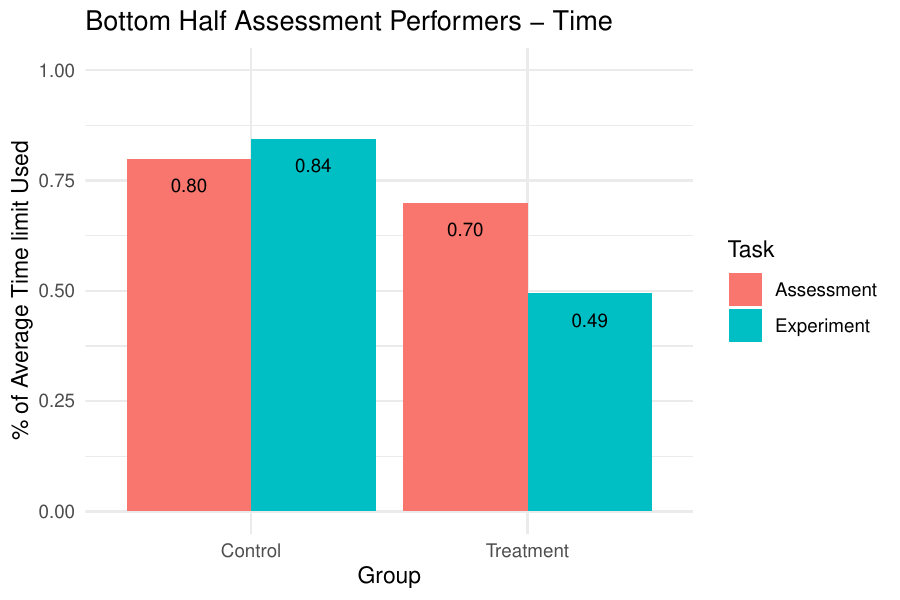}
\caption{Documents Task - Bottom-Half Assessment: Time}
\label{fig:bh_baseline_time}
\end{minipage}
\caption*{\textbf{Note:} The figures show the average performance on the Documents assessment task (red) and experiment task (green) (Figure \ref{fig:bh_baseline_score} - quality score, Figure \ref{fig:bh_baseline_time} - time) for the bottom-half of performers on the assessment task.}
\end{figure}

The analysis revealed that the most significant gains from using Gen AI were observed among participants in the lower 50\% of assessment task scores. These users, who initially demonstrated lower document comprehension skills, benefited most from the Gen AI tool. The average quality score for bottom-half participants that used Gen AI increased from 1.38 on the assessment task to 2.12 on the experiment task (see Figure \ref{fig:bh_baseline_score}). Time-to-completion for bottom-half participants decreased from 70\% of the allocated time limit on the assessment task to 49\% of the allocated time limit on the experiment task (see Figure \ref{fig:bh_baseline_time}). This is consistent with findings in \cite{NoyZhang2023} and \cite{Dellacqua:etal:2023}.

\begin{figure}[htbp]
\centering
\begin{minipage}{0.45\textwidth}
\centering
\includegraphics[width=\textwidth]{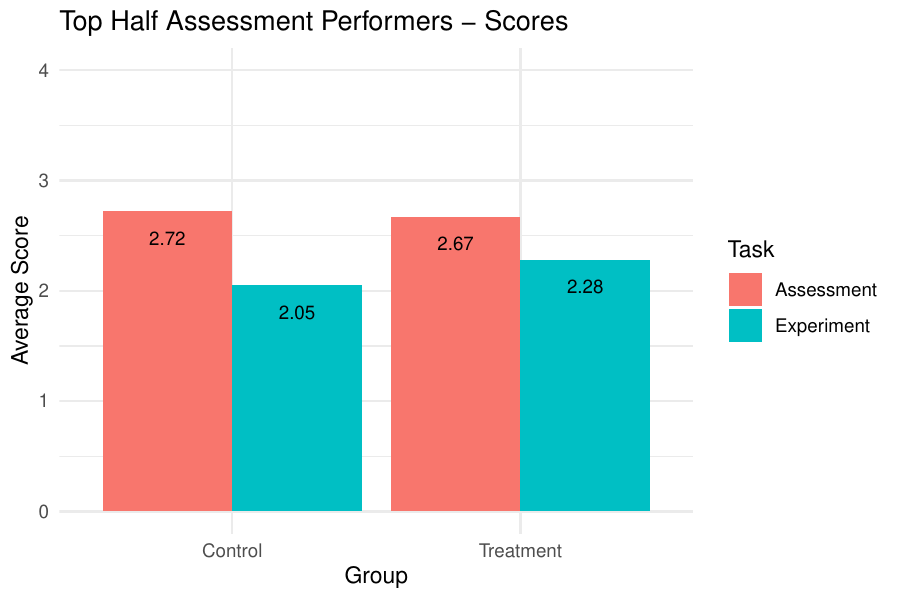}
\caption{Documents Task - Top-Half Assessment: Scores}
\label{fig:th_baseline_score}
\end{minipage}\hfill
\begin{minipage}{0.45\textwidth}
\centering
\includegraphics[width=\textwidth]{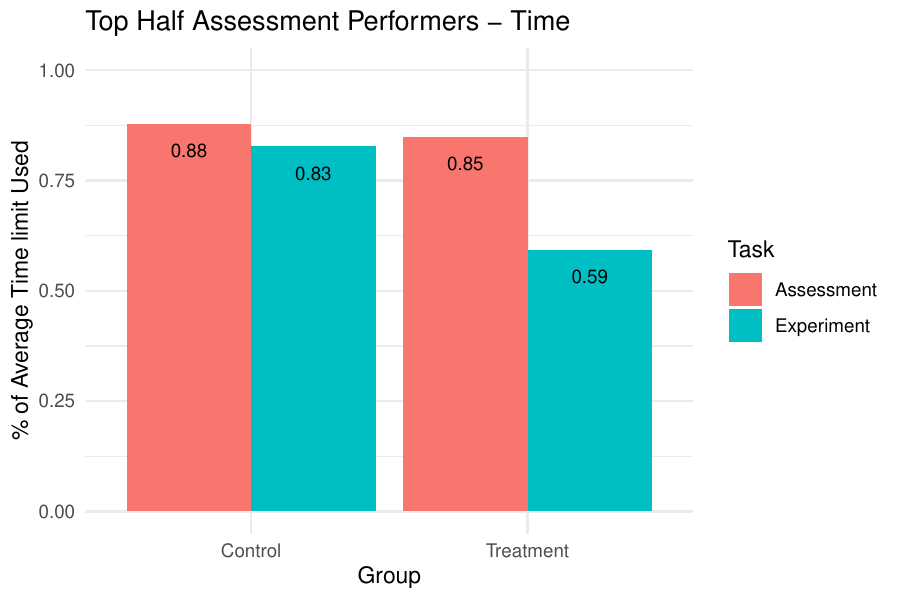}
\caption{Documents Task - Top-Half Assessment: Time}
\label{fig:th_baseline_time}
\end{minipage}
\caption*{\textbf{Note:} The figures show the average performance on the Documents assessment task (red) and experiment task (green) (Figure \ref{fig:th_baseline_score} - quality score, Figure \ref{fig:th_baseline_time} - time) for the top-half of performers on the assessment task.}
\end{figure}

For users in the top 50\% of baseline scores, Gen AI use on the experiment task led to a decrease in average quality score from 2.67 on the assessment task to 2.28 on the experiment task (see Figure \ref{fig:th_baseline_score}). This decline was less than that observed in the control group whose scores decreased from 2.72 on the assessment task to 2.05 on the assessment task. However, top-half Gen AI users also improved time-to-completion from 85\% of the allocated time limit on the assessment task to 59\% of the allocated time limit on the experiment task (see Figure \ref{fig:th_baseline_time}). This is similar to the effect in \cite{Dellacqua:etal:2023}. The decline in quality may reflect the higher baseline scores achieved on the shorter and slightly simpler assessment task. 

\subsection{Documents - Question Performance}

To further understand the efficacy of Gen AI in document comprehension tasks, we conducted an analysis of the responses generated for each individual question within the Documents task (detailed in Appendix \ref{sec:docsquestionperf}). The treatment group demonstrated strong performance on questions where the information required to answer the question was concentrated within one section of the document. Conversely, the treatment group performed more poorly on questions where a high quality answer required collating information from throughout the document. 

\subsection{Data - Primary Outcome: Quality}

The primary outcome of interest was the quality of responses to the Data task. The density plot (Figure \ref{fig:density_quality}) visualises the distribution of scores across the two groups and shows that the control group outperformed the treatment group with an average score of 36.5 out of 40 for control and 32.3 out of 40 for treatment. This equates to a 12\% reduction in quality for the treatment group. A two-tailed t-test confirmed that this difference is significant (p < 0.005).

Table \ref{dataregression} presents the results of the analyses examining the impact of using Gen AI on the quality outcome variable across the three model specifications. Column 1 uses only the treatment variable, while Columns 2 and 3 progressively incorporate additional covariates, including enrolment data and the baseline assessment. 

In column 1, the Gen AI treatment leads to a statistically significant decrease of 4.241 over the control, representing a 12\% reduction. In Column 2, after controlling for enrolment data covariates, the treatment effect remains negative but is lower in magnitude at 3.453. In Column 3, which includes both enrolment data and the baseline assessment, the treatment effect is a 3.860 reduction in quality and is again statistically significant (p < 0.05).

Across all specifications, the inclusion of additional covariates improves the model fit, as reflected in the increasing adjusted-R2 values (from 0.136 in Column 1 to 0.192 in Column 3). Model quality however decreases with the inclusion of these variables, with AIC increasing from 412.9 in Column 1 to 424.8 in Column 3, indicating poorer model performance.

Similar to the Documents task, as a robustness check, we use non-parametric estimators Targeted Maximum Likelihood and Generalised Random Forests, which again result in slightly larger effect size estimates compared to the linear model. These are shown with the associated standard errors in Table \ref{quality_effect_estimators_table_data}. The same set of covariates are included in both models as in Column 3 of Table \ref{dataregression}.

\begin{table}[h]  
    \centering  
    \caption{Data Task - Comparison of linear treatment effect with machine-learning estimators }  
    \label{quality_effect_estimators_table_data}  
    \begin{tabular}{lcc}  
        \toprule  
        Estimator & Effect Estimate & Standard Error \\  
        \midrule  
        Linear    & -3.860 & 1.419 \\  
        TMLE      & -4.302 & 1.255 \\  
        GRF       & -4.358 & 1.392 \\  
        \bottomrule  
    \end{tabular}  
\end{table}

\subsection{Data - Secondary Outcome: Task Completion Time}
The secondary outcome of interest was the time taken to complete the Data task. The results are visualised in the density plot (Figure \ref{fig:density_time}) and show that the average time for participants in the control group (32.5 minutes) was marginally lower when compared to the treatment group (33.5 minutes). This represents a 3\% increase in timing for the treatment group over the control group. However a two-tailed t-test confirms that this difference is not statistically significant (p-value = 0.7048).

\subsection{Data - Main Beneficiaries}

In contrast to the Documents task, on the Data task we did not observe a significant variation in performance difference from using Gen AI between the top and bottom assessment task performers. Our results showed that a majority of participants (n = 49) scored 100\% on the assessment task, with just 18 scoring less than 100\%. This indicates the assessment task was too simple to meaningfully differentiate the baseline competency of our participants.

\begin{figure}[htbp]
\centering
\begin{minipage}{0.45\textwidth}
\centering
\includegraphics[width=\textwidth]{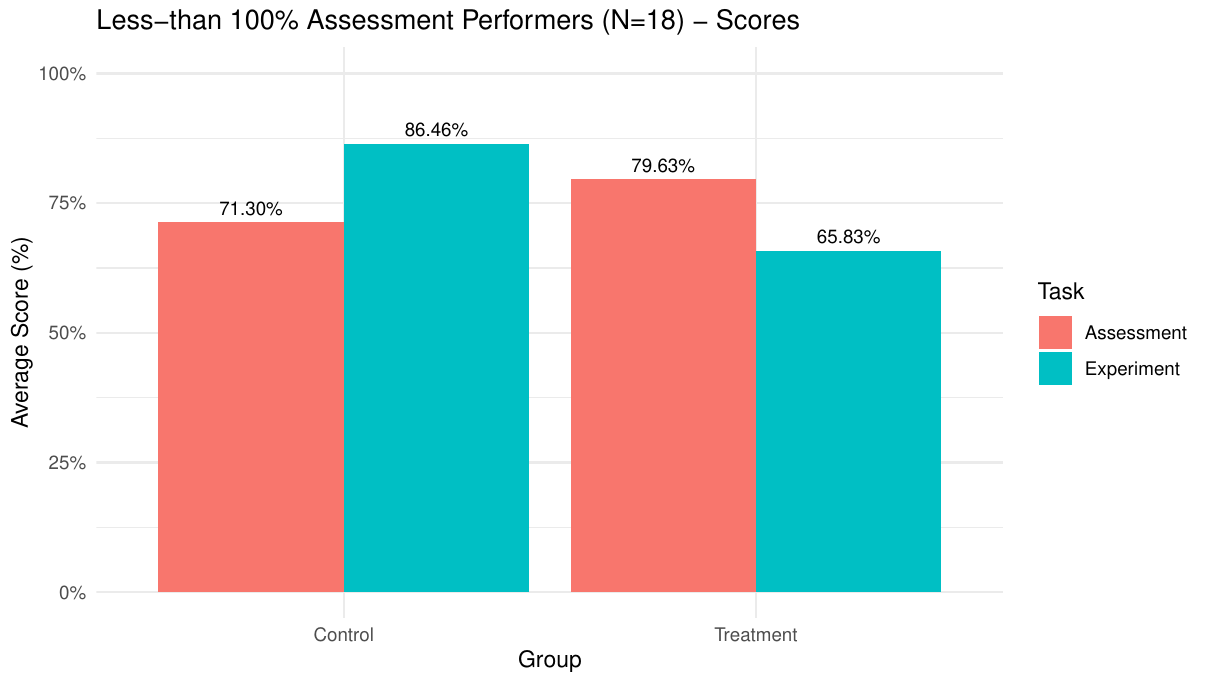}
\caption{Data Task - Less than 100\% on Assessment: Scores}
\label{fig:data_bh_baseline_score}
\end{minipage}\hfill
\begin{minipage}{0.45\textwidth}
\centering
\includegraphics[width=\textwidth]{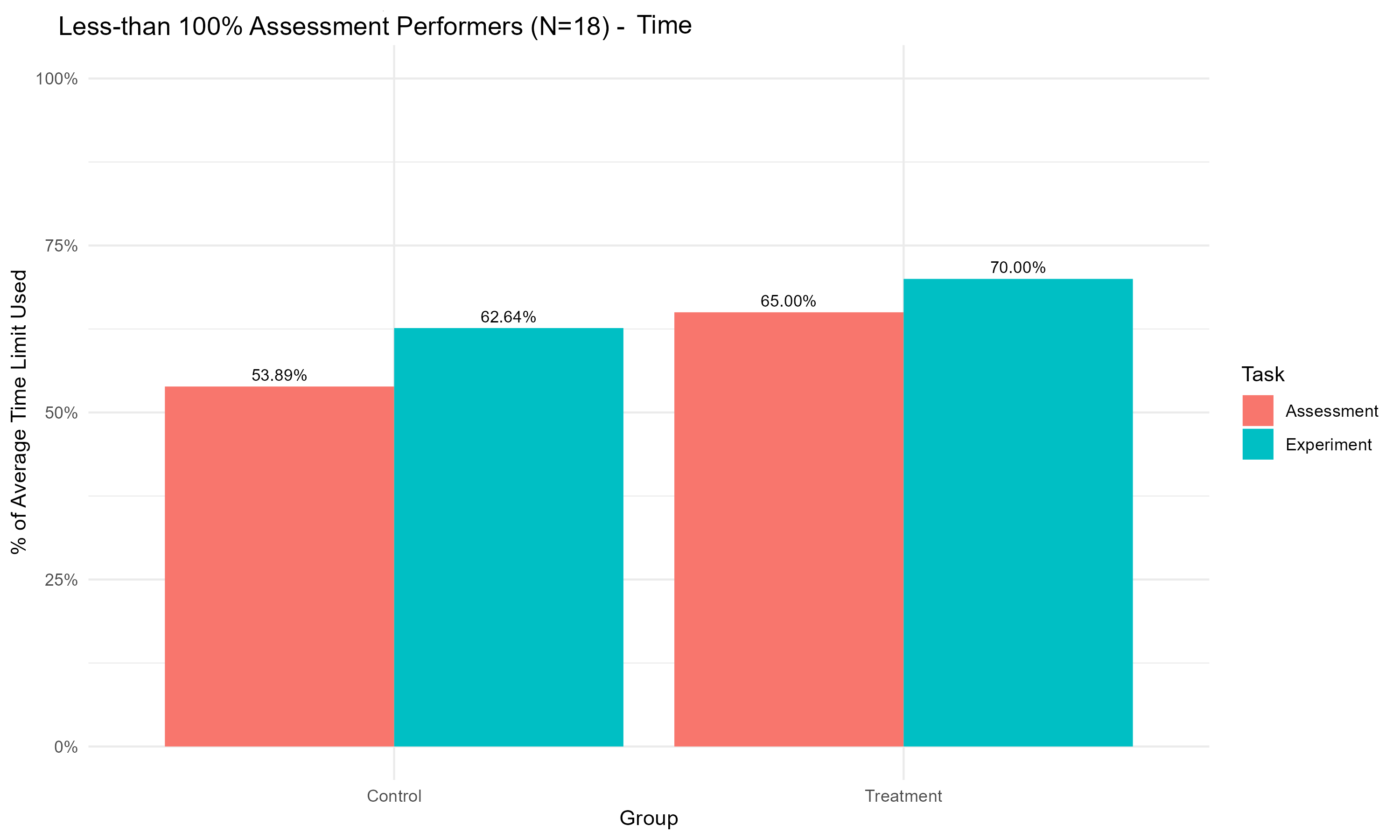}
\caption{Data Task - Less than 100\% on Assessment: Time}
\label{fig:data_bh_baseline_time}
\end{minipage}
\caption*{\textbf{Note:} The figures show the average performance on the Data assessment task (red) and experiment task (green) (Figure \ref{fig:data_bh_baseline_score} - quality score, Figure \ref{fig:data_bh_baseline_time} - time) for those participants who scored less than 100\% on the assessment task.}
\end{figure}

For those who scored less than 100\% on the assessment task, Gen AI use on the experiment task led to a decrease in percentage of correct answers from 79.63\% on the assessment task to 65.83\% on the experiment task (see Figure \ref{fig:data_bh_baseline_score}). Time-to-completion for those who scored less than 100\% on the assessment task increased from 65\% of the allocated time limit on the assessment task to 70\% of the allocated time limit on the experiment task (see Figure \ref{fig:data_bh_baseline_time}). 

\begin{figure}[htbp]
\centering
\begin{minipage}{0.45\textwidth}
\centering
\includegraphics[width=\textwidth]{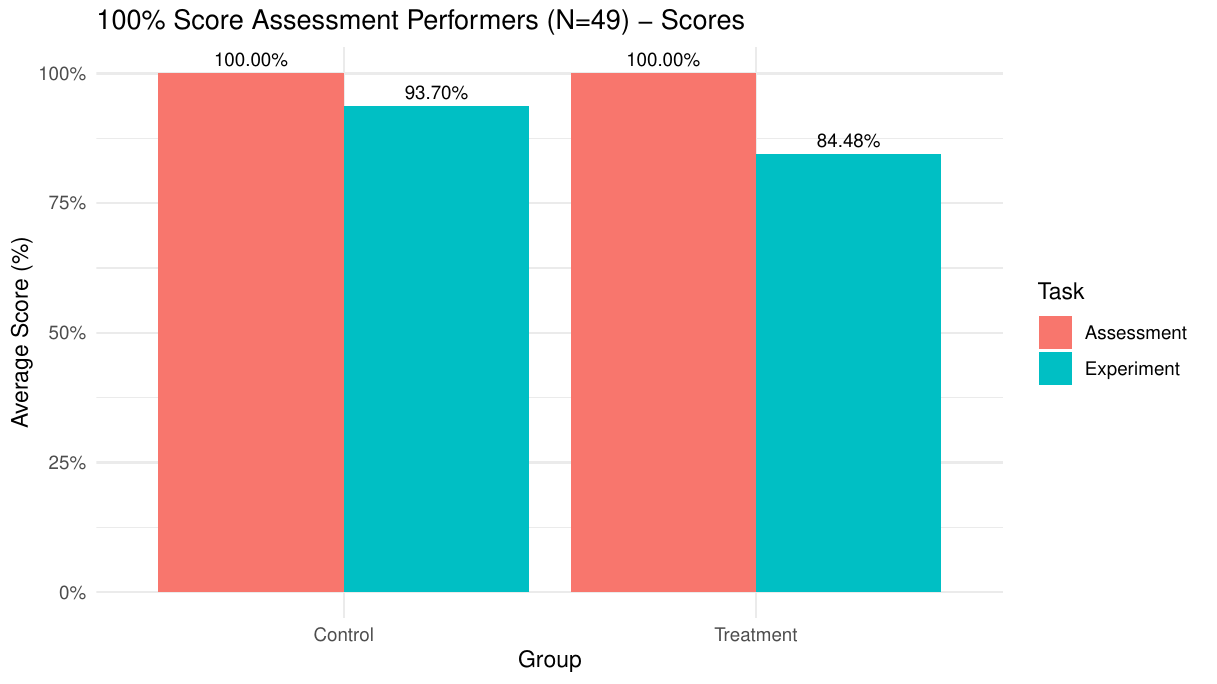}
\caption{Data Task - 100\% on Assessment: Scores}
\label{fig:data_th_baseline_score}
\end{minipage}\hfill
\begin{minipage}{0.45\textwidth}
\centering
\includegraphics[width=\textwidth]{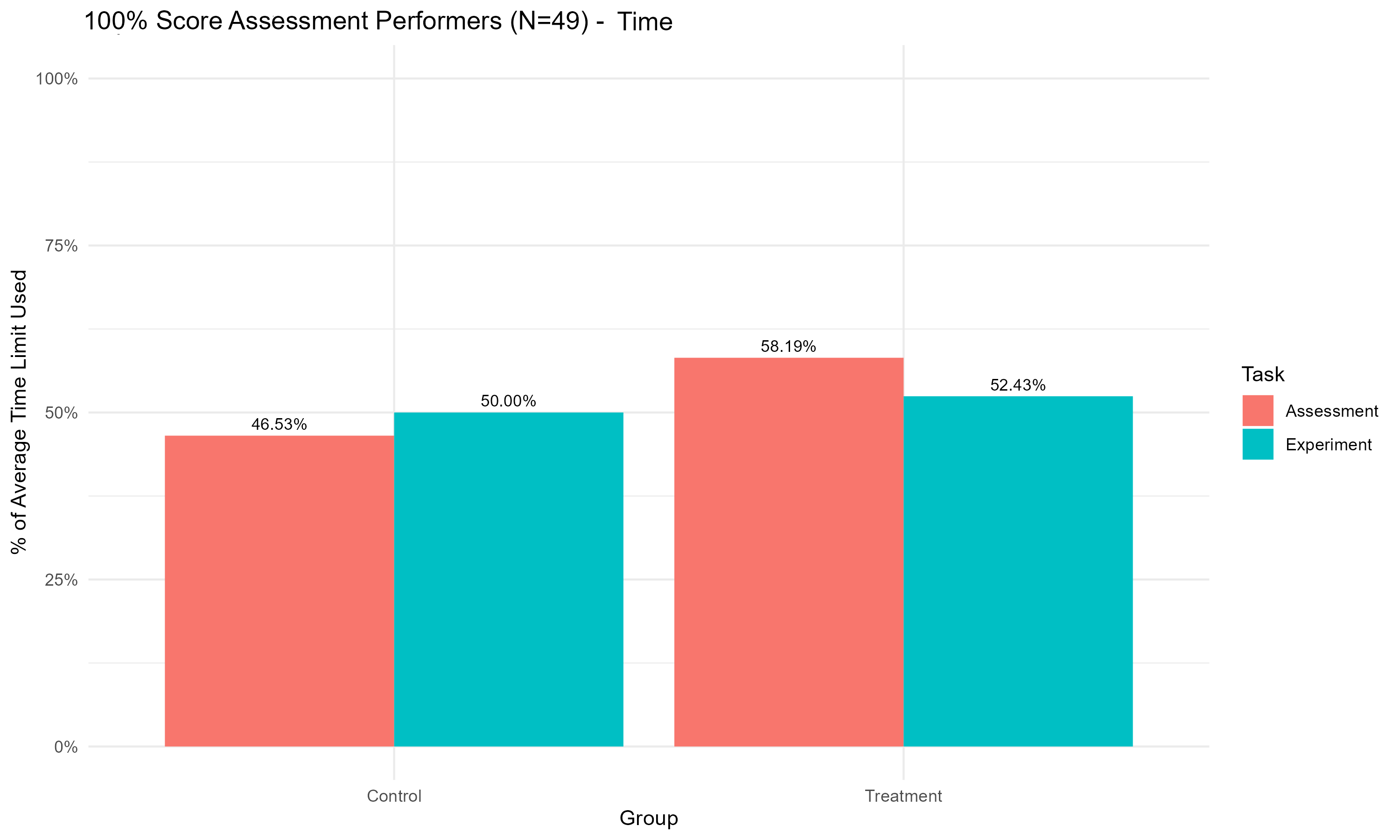}
\caption{Data Task - 100\% on Assessment: Time}
\label{fig:data_th_baseline_time}
\end{minipage}
\caption*{\textbf{Note:} The figures show the average performance on the Data assessment task (red) and experiment task (green) (Figure \ref{fig:data_th_baseline_score} - quality score, Figure \ref{fig:data_th_baseline_time} - time) for those participants who scored 100\% on the assessment task.}
\end{figure}

For those who scored 100\% on the assessment task, Gen AI use on the experiment task led to a decrease in percentage of correct answers from 100\% to 84.48\% (see Figure \ref{fig:data_th_baseline_score}). Time-to-completion for those who scored 100\% on the assessment task decreased from 58.19\% of the allocated time limit on the assessment task to 52.43\% of the allocated time limit on the experiment task (see Figure \ref{fig:data_th_baseline_time}).

\subsection{Results Discussion}

The improvements in quality and speed shown in the Documents task indicate that Gen AI can play a valuable role in augmenting human performance on information processing tasks in a public sector context. The observed quality improvement for the Documents task is slightly lower than the quality improvement in \cite{NoyZhang2023} and significantly lower than reported in \cite{Dellacqua:etal:2023}, while the improvement in speed is within the range reported in those two studies. 

The high performance on simpler questions of the treatment group using Gen AI in the Documents task demonstrates the potential of Gen AI to streamline certain types of queries, offering valuable support in retrieving targeted information efficiently. However, the poorer performance on more complex questions underscores the current limitations when collating scattered information, reinforcing the need for human oversight.

\begin{figure}[htbp]
\centering
\begin{minipage}{0.5\textwidth}
\centering
\includegraphics[width=\textwidth]{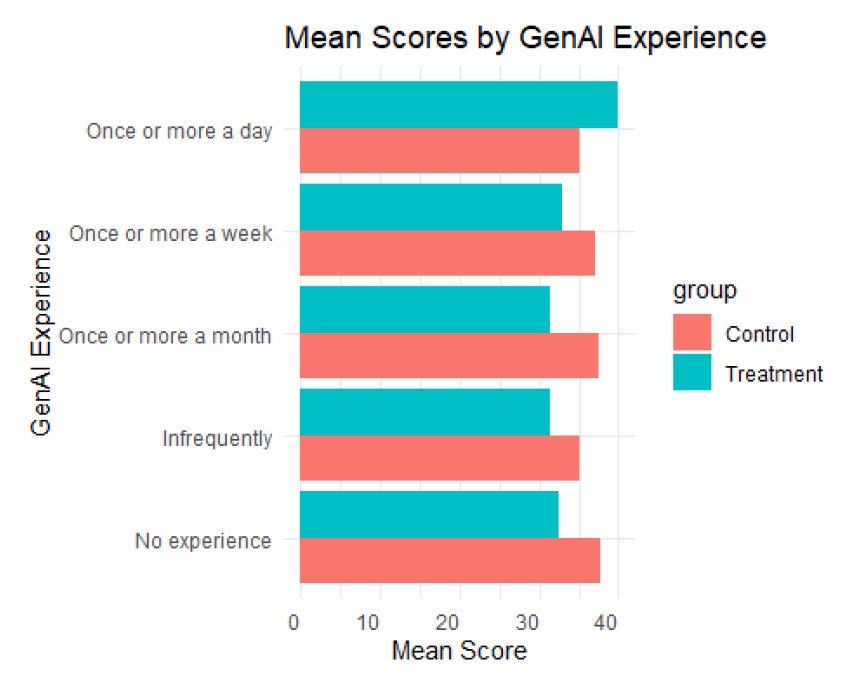}
\caption{Data Task - Mean Quality Scores by Gen AI Experience}
\label{fig:genai_experience_quality}
\end{minipage}\hfill
\caption*{\textbf{Note:} The figure shows the average quality score on the Data experiment task for control (red) and treatment (green) groups, split by Gen AI experience levels from the enrolment survey.}
\end{figure}

For the Data task, the non-randomisation across tasks meant that staff with data skills/roles self-selected into this task. This is perhaps one of the reasons for the observed results, which we discuss further in section \ref{sec:lessonslearned}.
In the Data task, those in the treatment group with more experience using Gen AI had higher quality scores compared to those with less experience (Figure \ref{fig:genai_experience_quality}). This suggests that prior exposure to Gen AI may have resulted in improved performance on the task when using the Gen AI application.

\section{Field Notes and Lessons Learned}
\label{sec:lessonslearned}

This section contains some additional detail that we are including to explain the rationale for our design and that may assist other public sector organisations in conducting their own trials. These include areas where we know the trial could be improved upon.  

\subsection{Design}

\subsubsection{Randomisation Across Tasks}

An open invitation was issued to all staff at CBI to participate in the trial tasks through an internal website and in-person briefings to individual business areas.

On registration, participants were asked to specify a preference for one of the two tasks (Documents or Data). In order to prevent attrition, participants were largely assigned to a task based on the preference they had indicated. This resulted in selection bias, in that the Data task comprised a higher proportion of technical and data professionals than the general population of the CBI. This may be the cause of the relatively high scores of those in the control group on the Data task, as they had a high baseline competency.

During the conceptual work on the design phase, we explicitly discussed randomising across tasks but were concerned that it would result in more attrition as we did not have similar baseline to judge the risk of one downside (sample selection into the Data task) versus another (attrition risk). We would randomise a larger sample of participants across tasks in future trials to ensure a more representative sample in each task.

\subsubsection{Attrition}
Both trials lost around 25\% of participants to attrition. This was due to a combination of factors. First, we could not incentivise completion of the trial through, for example, financial recognition, which may have resulted in a higher completion rate. Second, these included the unavailability of some participants due to holidays during the experiment window, and time constraints with participants' existing work commitments. In future trials, we would recruit a larger population to deal with attrition.

\subsubsection{Question phrasing}

In the Documents task, a number of questions were phrased in a way that they could be reasonably answered with simply "yes" or "no" (the full list of questions for the task is in Appendix \ref{sec:documentstasks}). This was particularly evident for questions 3 ("Is the firm’s organisational structure clearly documented?") and 4 ("Are roles and responsibilities clearly identified in the Board and sub-committees?") where 35\% and 25\% of answers respectively from the control group were variations of "yes" or "no". This compares to <5\% of "yes"/"no" answers from the treatment group for these questions. These "yes"/"no" answers scored a 0 or 1 out of a maximum score of 4 based on the evaluation rubric, whereas a more detailed answer with supporting context scored higher. The Gen AI application used by the treatment group generally provided a more detailed response, so this may have impacted the quality scores across the treatments for these questions. In future trials we would phrase questions in a way that makes clear that a better quality answer should include additional detail.

\subsubsection{Time: a self-reported secondary outcome}
Participants were provided with instructions and carried out the experiment in their own time. They were instructed to adhere to a time limit, but we had no way to enforce this. We could not easily utilise a social science experiment platform (e.g. Gorilla) to manage the tasks as this would have involved incorporating a Gen AI application into a platform. Time values were therefore self-reported by each participant. In future trials we would look to integrate applications into an experiment platform to ensure greater control over the trial.

\subsubsection{Additional treatment groups}
In the design, we considered creating two treatment groups - one group who completed the task with the aid of a Gen AI application, and one group who received prompt training in addition to access to the Gen AI application to assess the differential effect of prompt training on the outcome variables. Ultimately we decided against this given our sample size. With a larger sample size, we would consider prompt training, including the importance of question phrasing and popular prompting techniques as an additional treatment. 

\subsection{Execution}

\subsubsection{Answer Evaluation}
The Documents task involved a human evaluation panel (consisting of subject matter experts within the CBI) scoring participants' responses based on an evaluation rubric. Each response was scored by two evaluators. This was a time consuming manual task, requiring several hours from each evaluator. An opportunity to improve the efficiency of the evaluation process would be to utilise an automated labelling platform or to use Gen AI to evaluate responses by incorporating the evaluation rubric. This would significantly reduce the time taken to evaluate the quality in such experiments.

\subsubsection{IT Operations}
During testing, several users had access issues accessing the Gen AI applications used in the trials. These were resolved, but delayed the start of the trial. There were intermittent unexplained outages (of up to 3 hours) of the applications during the one week experiment window. We do not know the effects of these outages on the results.

\subsection{Feedback}

The quantitative results above followed the statistical analysis plan in the pre-registered experiment. We conducted two supplementary pieces of analysis not included in the statistical analysis plan at registration in order to gain qualitative insights from participants - a results feedback workshop and a post-trial survey.

The feedback workshop involved communicating preliminary trial results to participants and gathering qualitative feedback. Approximately 100 participants attended this workshop.

Following trial completion, we issued a survey to the 175 participants that completed at least the assessment phase of the trials to gather feedback on their experience. We received responses from 51 participants, a 29\% response rate.

For those in the treatment group, 66\% found the application to be very effective or effective in helping them complete the task. This illustrates a generally positive attitute from participants towards the use of Gen AI for completing the tasks.

The survey further evaluated participants' use of the Gen AI application to complete the tasks. For those in the treatment group, just 31\% of respondents copied the questions from the task directly into the Gen AI application prompt, while 69\% either shortened or rephrased the question before entering the prompt. In terms of outputs, just 33\% of respondents took the responses directly from the Gen AI application as their final answer, while 50\% used a combination of the Gen AI response and their own knowledge to formulate their answer. This aligns with the ``Cyborg'' style of human-AI interaction noted by \cite{Dellacqua:etal:2023}, where users fully integrate AI and human capabilities into their work.

Another finding from the treatment group was the expressed interest in experimenting with the Gen AI tool and verifying the results it generated. Rather than focusing on completing the task in the shortest time, many participants engaged in exploring the tool's capabilities. They reported a tendency to cross-check the Gen AI tool's responses against their own interpretations of the document, prioritising the accuracy and reliability of the output. This behaviour suggests that the recorded time to completion for the Documents task, while significantly improved for the treatment group, may be understated. Participants were not solely aiming to finish the task as quickly as possible but were instead iteratively refining their queries and reviewing the tool's results. This indicates a learning process and a focus towards understanding the tool's output rather than simply relying on it for rapid completion.

Several participants indicated they would like further training on effective prompting, and that further use of Gen AI tools would be beneficial to improve their competency. With more structured instruction on how to craft effective prompts and refine the outputs, it is plausible that the overall quality of the task responses could have been further improved. 

Some participants noted they were not sure whether to trust the response produced by the Gen AI application. A potential enhancement for future Gen AI applications would be to display the model's confidence or probability in the response to the user, which would help build user trust.

\section{Conclusion}
In this paper, we study the effectiveness of Gen AI on two types of knowledge-based tasks in a public sector context - obtaining information from complex documents, and analysing data. We used a randomised control trial that showed that Gen AI can provide significant quality and efficiency benefits, but that it depends on the task.

For the Documents task, the use of Gen AI resulted in an improvment in quality and speed. Quality improved by 17\% while speed improved by 34\% for the group using Gen AI versus a control group. Gen AI also proved to benefit lower-half performers the most, though even top-half performers saw significant speed improvements from using Gen AI. Conversely, for the Data task, the use of Gen AI resulted in a 12\% reduction in quality, and no signficant difference in speed.

Our findings have implications for the use of Gen AI in a public sector context. Importantly, the results suggest that adoption of Gen AI in the public sector should be task-dependent, and organisations must consider carefully where to integrate Gen AI into workflows to ensure that the optimal balance of human knowledge and Gen AI is achieved. Document comprehension tasks, which often require synthesizing information from well-structured texts, are more amenable to Gen AI assistance. In contrast, data analytics tasks may demand a higher level of precision and domain-specific expertise that Gen AI tools have yet to consistently achieve. Additionally, the self-selection bias observed in the Data task -- where more technically proficient participants opted for data-related tasks -- may have influenced the overall performance outcomes. 

Our results show that within tasks, the use of Gen AI resulted in varying performance. On the Documents task, questions involving complex concepts resulted in poorer performance for the Gen AI group. This suggest more complex tasks require better context to be provided to a Gen AI tool, and highlights the need for human review of Gen AI outputs. This will require public sector knowledge workers to incorporate Gen AI as a tool in their workflows, rather than simply using its outputs verbatim. Indeed, in our study, a majority of participants reported editing Gen AI outputs based on their own knowledge to formulate an answer.

Operational challenges encountered during the trial, such as technical issues with accessing Gen AI applications and the lack of structured training for effective AI utilization, further highlight areas for improvement. Participant feedback indicated a desire for more comprehensive training on prompt engineering and a need for greater trust in AI-generated outputs, underscoring the importance of human-AI collaboration rather than reliance on AI as a standalone solution. Despite these limitations, this study underscores the substantial potential of Gen AI to enhance quality and efficiency in public sector knowledge work, particularly in tasks that benefit from rapid information processing and summarization. However, it also cautions against overreliance on Gen AI for complex analytical tasks where human expertise remains paramount. Future research should aim to expand the sample size and diversify the range of tasks to better understand the broader applicability of Gen AI in various public sector contexts. Additionally, improving Gen AI tools to better handle complex data analytics and integrating structured training programs for users can enhance the effectiveness and reliability of AI assistance. Longitudinal studies could also explore the long-term impacts of Gen AI adoption on public sector workflows and decision-making processes.

Our study provides evidence for the benefits of Gen AI on public sector knowledge work. While Gen AI holds promise as a transformative tool for enhancing public sector effectiveness and efficiency, its successful implementation requires careful consideration of task suitability, user training, and continuous evaluation to ensure that it enhances the quality and integrity of public services.

\section*{Acknowledgements}

The Authors wish to acknowledge the contribution of their colleagues: Tom Cleary, Niamh O’Malley, Alec Stone, Gerry Langtry, Wael Muhammad, David Ring, Edward Roche, Deirdre Howard, Ian Goff, Liam Fallon, John Fitzgerald, Emma Wall, Mayuri Sulibhavi, Anthony Hayes, Patrick Clarke, Mark Heaney, Gavin Russell, and all participants in the trials.

\newpage
\appendix
\section{Appendix - Experiment Design}
\label{sec:appendixtasks}
This appendix provides additional details on the enrolment survey, and the tasks carried out by participants during the trials.

\subsection{Enrolment survey}
\label{sec:enrolmentsurvey}

We conducted an enrolment survey to gather various data on participants, including job information (role, division, tenure), demographic information (gender, educational background), tech openness, and previous experience with Gen AI. We used these to assess the representativeness of our simple random assignment of participants to treatment and control groups, and as control variables in our regression models.

\subsection{Tasks}

Both experiments started with an assessment task that all participants completed without the aid of Gen AI, in order to assess their baseline competency in completing these types of tasks and allow us to control for this in our regression models. Participants were then divided into treatment and control groups, and completed the experiment task, either with or without the aid of Gen AI according to their treatment assignment. The tasks are described below.

\subsubsection{Documents}
\label{sec:documentstasks}

\paragraph{Assessment Task}

The assessment task was based on a 2022 Solvency and Financial Conditions Report for an Irish insurer, an approximately 100 page document containing information on the firm's business and performance, system of governance, risk profile, valuation for solvency purposes and capital management process.

Participants were asked to complete four questions (below) that required them to review and extract information from the document in order to formulate their answers. 

\begin{enumerate}
\item What types of business does the firm currently write and how profitable have these lines been for 2022 and 2021?
\item How is the firm addressing potential emerging risks, if at all?
\item Assess the firm's recent business performance.
\item What are the key responsibilities of the Chairman and Executive Management Team?
\end{enumerate}

Participants were instructed to adhere to a 30 minute time limit, and submit responses via an online survey tool.

Responses were evaluated by independent human graders using an evaluation rubric. An example of the rubric for one question is provided in Table \ref{tab:assessmentrubric}. Each response was graded on a scale from 0 (incorrect) to 4 (very good).

\newpage
\small
\begin{tabularx}{\textwidth}{X|p{0.7cm}|p{7cm}}
\hline
Question&Grade&Example\\
\hline
How is the firm addressing potential emerging risks, if at all?&4&An emerging risk is a risk which may or may not develop, is difficult to quantify, may have a high loss potential 
and is marked by a high degree of uncertainty. We have a defined process in place for the identification of and response to emerging risks, which is informed through the use of subject matter experts, workshops, Risk and 
Control Self Assessments and consulting a range of external resources. Key emerging risks are monitored regularly by the Board and Risk Committees to assess whether they might become significant for the business  and require specified action to be taken. Key Emerging Risks include:
\begin{itemize}[topsep=0pt,itemsep=0pt]
\item An increased frequency of cyber attacks, and the impact that these factors may have on society’s future 
insurance needs and claims types and frequencies.
\item Stricter Emission Targets and ESG driven market change.
\item Technological advances changing the shape of the insurance industry and competitive environment.
\item Changes in customer behaviour including the potential expectation to communicate largely through 
mobile channels or the expectation of self-service and self-solve.
\item Global deterioration in economic conditions and particularly in Ireland may lead to a reduction in revenue 
and profits.
\item Global socio-political uncertainty that may cause an adverse impact on profitability.
\item Evolving regulatory and legislative landscape. We continuously monitor developments at both a local and 
EU level to ensure continued compliance with legislative and regulatory requirements.
\item Availability of reinsurance may be restricted due to macroeconomic, environmental and/or market 
developments
\end{itemize}
\\
\hline
How is the firm addressing potential emerging risks, if at all?&3&We have a defined process in place for the identification of and response to emerging risks, which is informed through the use of subject matter experts, workshops, Risk and 
Control Self Assessments and consulting a range of external resources. Key Emerging Risks include:
\begin{itemize}[topsep=0pt]
\item An increased frequency of cyber attacks, and the impact that these factors may have on society’s future 
insurance needs and claims types and frequencies.
\item Stricter Emission Targets and ESG driven market change.
\item Technological advances changing the shape of the insurance industry and competitive environment.
\item Changes in customer behaviour including the potential expectation to communicate largely through 
mobile channels or the expectation of self-service and self-solve.
\end{itemize}
\\
\hline
How is the firm addressing potential emerging risks, if at all?&2&Key Emerging Risks include:
\begin{itemize}[topsep=0pt]
\item An increased frequency of cyber attacks, and the impact that these factors may have on society’s future 
insurance needs and claims types and frequencies.
\item Stricter Emission Targets and ESG driven market change.
\item Technological advances changing the shape of the insurance industry and competitive environment.
\item Changes in customer behaviour including the potential expectation to communicate largely through 
mobile channels or the expectation of self-service and self-solve.
\end{itemize}
\\
\hline
How is the firm addressing potential emerging risks, if at all?&1&Some Key Emerging Risks are addressed in the SFCR.
\\
\hline
How is the firm addressing potential emerging risks, if at all?&0&The firm does not look at emerging risks in its SFCR.
\\
\hline
 \caption{Documents - assessment task - evaluation rubric example}
 \label{tab:assessmentrubric}
\end{tabularx}

\paragraph{Experiment Task}

Similar to the assessment task, the experiment task was based on a Solvency and Financial Conditions Report for another Irish insurer.

Participants were asked to complete twelve questions (below) that required them to review and extract information from the document in order to formulate their answers. 

\begin{enumerate}
\item Have there been any changes in the composition, credit rating, duration or liquidity of the investment portfolio?
\item Describe the composition of the investment portfolio of the firm.
\item Is the firm’s organisational structure clearly documented?
\item Are roles and responsibilities clearly identified in the Board and sub-committees?
\item Has there been changes to the Capital Risk Appetite or Strategic Solvency Target and rationale?
\item Is there adequate reporting of the internal control framework?
\item Describe the firm’s outsourcing arrangements.
\item Describe the governance arrangements around outsourcing within the firm.
\item Provide a succinct description of the firm's Business Model.
\item What entity is the direct shareholder of the firm and what entitiy is the ultimate shareholder of the firm? 
\item Describe the Fit \& Proper processes and Procedures within the firm?
\item How has the firm's underwriting performed in 2022 compare to 2021?  
\end{enumerate}

Participants were instructed to adhere to a 60 minute time limit, and submit responses via an online survey tool.

Responses were evaluated by independent human graders using an evaluation rubric. An example of the rubric for one question is provided in Table \ref{tab:experimentrubric}. Each response was graded on a scale from 0 (incorrect) to 4 (very good).

\newpage
\small
\begin{tabularx}{\textwidth}{X|p{0.7cm}|p{7cm}}
\hline
Question&Grade&Example\\
\hline
Provide a succinct description of the firm's Business Model.&4&The firm is a non-life insurance company. The principal activity of the Company is the transaction of property, motor, liability and marine insurance business
within the Republic of Ireland. The Company offers a wide range of non-life insurance products to both retail and corporate customers. The Company is one of the leading non-life insurers in Ireland with Gross premium written of X euro. The main risks and therefore drivers of the risk capital requirement are insurance related risks. The Company’s strategic position is to maintain and expand its underwriting portfolio through accepting exposures at economic prices and providing good value and quality service to its policyholders.
\\
\hline
Provide a succinct description of the firm's Business Model.&3&The firm is a non-life insurance company  whose  principal activity is the transaction of property, motor, liability and marine insurance business. The Company offers a wide range of non-life insurance products to both retail and corporate customers. The Company is one of the leading non-life insurers in Ireland with Gross premium written of X euro. The Company’s strategic position is to maintain and expand its underwriting portfolio through accepting exposures at economic prices and providing good value and quality service to its policyholders.
\\
\hline
Provide a succinct description of the firm's Business Model.&2&The firm is a non-life insurance company  whose  principal activity is the transaction of property, motor, liability and marine insurance business. The Company offers a wide range of non-life insurance products to both retail and corporate customers.
\\
\hline
Provide a succinct description of the firm's Business Model.&1&The firm is a non-life carrier writing multiple lines of business in ROI.
\\
\hline
Provide a succinct description of the firm's Business Model.&0&There is no description of the firm's business model in the document.
\\
\hline
 \caption{Documents - experiment task - evaluation rubric example}
 \label{tab:experimentrubric}
\end{tabularx}

\subsubsection{Data}
\label{sec:datatasks}

\paragraph{Assessment Task}

The assessment task was based on the Quantitative Reporting Templates (QRTs) between 2019 and 2022 for an Irish insurer. In the assessment task, participants were asked to consider four questions (listed below) by reviewing the QRT data and populating the answers to the questions in the table provided for each question in the answersheet. Participants were also provided with an 'Assistance Document' which explained the relevant terms and data points needed. They were allocated a period of 30 minutes in which to complete the task and were asked to include the completion time at the end of the answer sheet. 

\begin{enumerate}
\item What is the total assets of the firm for 2021 and 2021?
\item What is the direct Gross Written Premium (GWP) for the firm by country for 2020 for the countries Spain, France and Great Britain?
\item Did equities increase value for the firm from 2020 to 2021?
\item Has there been a deterioration in the claims loss ratio for the firm for income protection insurance from 2020 to 2021?
\end{enumerate}

To grade the submitted answers two individuals with the required domain knowledge relating to QRTs determined the correct answers for each question in the assessment task. In the grading of the responses, answers were deemed to be either fully correct (and allocated a score of 1) or incorrect (and allocated a score of 0). In total, across the four questions, there were nine different values to retrieve and participants were assigned a final score between 0 and 9.

\paragraph{Experiment Task}

The experiment task was based on the Quantitative Reporting Templates (QRTs) between 2019 and 2022 for another Irish insurer. In the experiment task, participants were asked to consider six questions (listed below) by reviewing the QRT data and populating the answers to the questions in the table provided for each question in the answersheet. Participants were also provided with an 'Assistance Document' which explained the relevant terms and data points needed. They were allocated a period of 60 minutes in which to complete the task and were asked to include the completion time at the end of the answer sheet. 

\begin{enumerate}
\item Has the Solvency Capital Requirement (SCR) ratio for the firm changed from 2021 to 2022?
\item In 2022, what were the euro amounts of the components of the SCR for the firm?
\item What were the Tiers of Total eligible Own Funds to meet the SCR from 2021 to 2022 for the firm and how have they changed?
\item What is the breakdown of direct business Gross Written Premium (GWP) by SII line of business in 2022 for the firm?
\item What was the Gross Combined Operating Ratio for direct motor vehicle liability insurance for the firm in 2022?
\item Provide a breakdown of value of the categories of bonds listed (in euro) on the SII balance sheet for the firm in 2022.
\end{enumerate}

To grade the submitted answers two individuals with the required domain knowledge relating to QRTs determined the correct answers for each question in the trial task. In the grading of the responses, answers were deemed to be either fully correct (and allocated a score of 1) or incorrect (and allocated a score of 0). In total, across the six questions, there were forty different values to retrieve and participants were assigned a final score between 0 and 40.

\newpage
\section{Appendix - Analysis of Question Performance}

\subsection{Documents}
\label{sec:docsquestionperf}

To further understand the efficacy of Gen AI in document comprehension tasks, we conducted an analysis of the responses generated for each individual question within the Documents task. This analysis revealed some patterns in the treatment group's performance, indicating areas where answers provided with assistance from Gen AI were judged to be high quality by human evaluators versus where quality was judged to be lower.

\begin{figure}[htbp]
\centering
\begin{minipage}{0.7\textwidth}
\centering
\includegraphics[width=\textwidth]{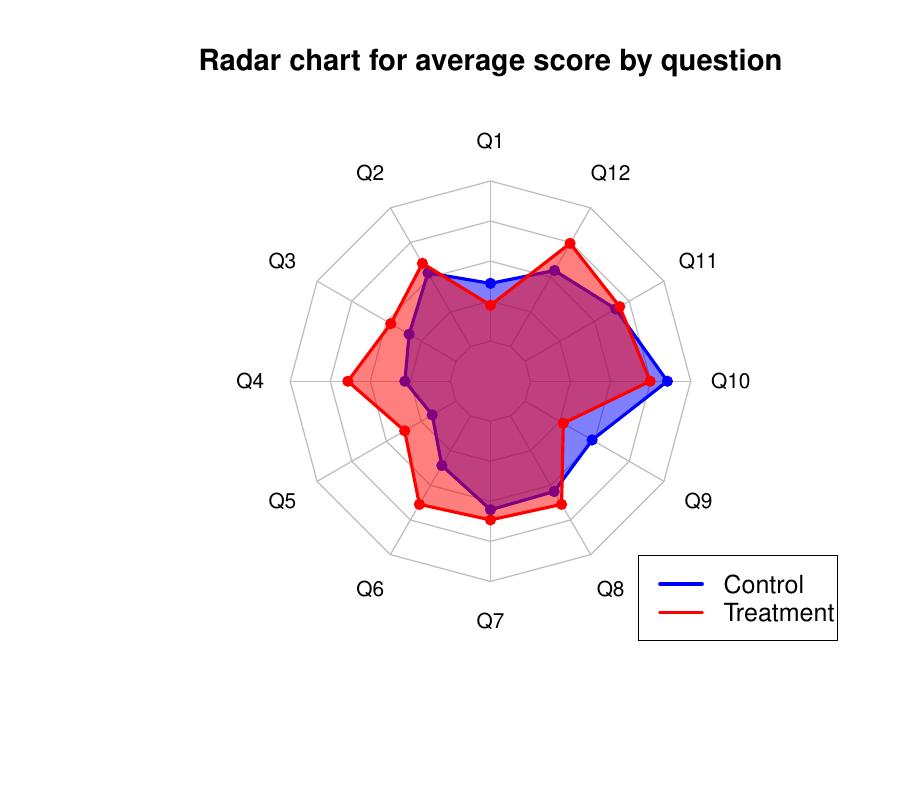}
\caption{Average Scores by Question}
\label{fig:radar_chart}
\end{minipage}\hfill
\end{figure}

\begin{figure}[htbp]
\centering
\begin{minipage}{0.45\textwidth}
\centering
\includegraphics[width=\textwidth]{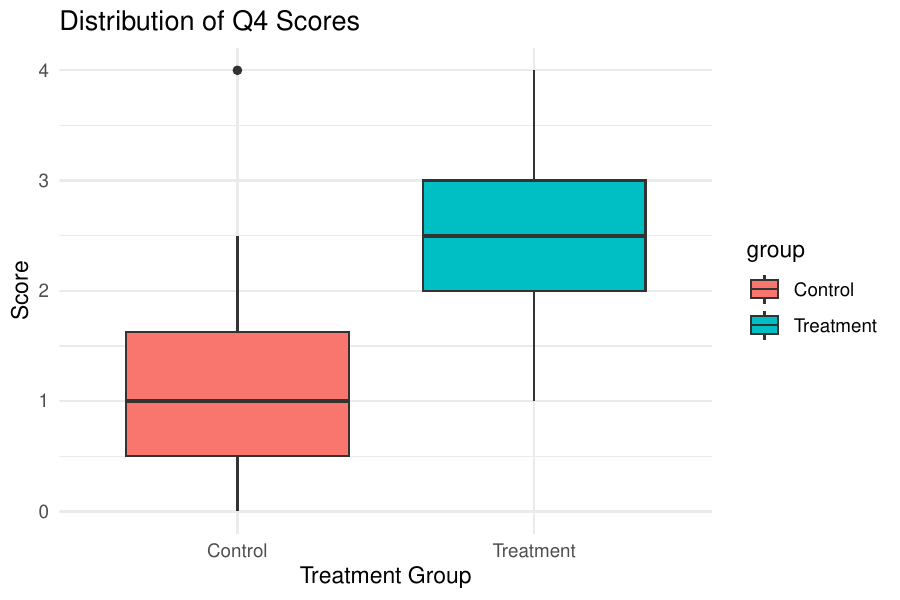}
\caption{Q4 Boxplot}
\label{fig:q4_boxplot}
\end{minipage}\hfill
\begin{minipage}{0.45\textwidth}
\centering
\includegraphics[width=\textwidth]{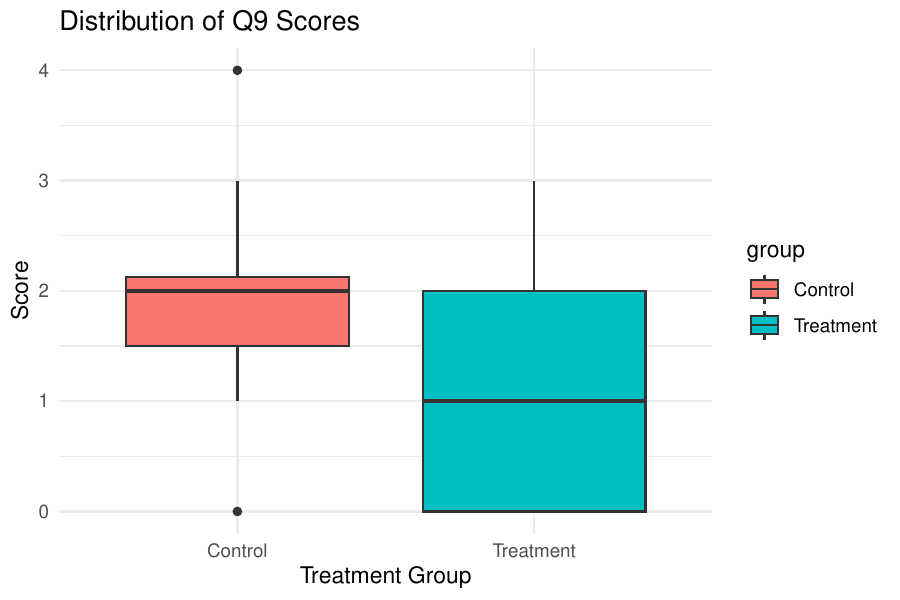}
\caption{Q9 Boxplot}
\label{fig:q9_boxplot}
\end{minipage}
\end{figure}

The treatment group demonstrated strong performance on questions where the information required to answer the question was concentrated within one section of the document. Examples of this are evident in Figure \ref{fig:radar_chart}. Question 4 (``Are roles and responsibilities clearly identified in the Board and sub-committees?'') is a question where a high quality answer required sourcing information from one section of the document. The control group obtained an average quality score of 1.14 in this question compared to 2.56 for the treatment group (see Figure \ref{fig:q4_boxplot}), indicating that Gen AI was able to accurately identify the information required to answer this question and significantly improve the quality of responses. Conversely, the treatment group performed more poorly (see Figure \ref{fig:q9_boxplot}) on question 9 (``Provide a succinct description of the firm's Business Model?''). In this question, where a high quality answer required collating information from throughout the document, the control group outperformed the treatment group with an average score of 1.93 compared to 1.1.

\begin{figure}[htbp]
\centering
\begin{minipage}{0.42\textwidth}
\centering
\includegraphics[width=\textwidth]{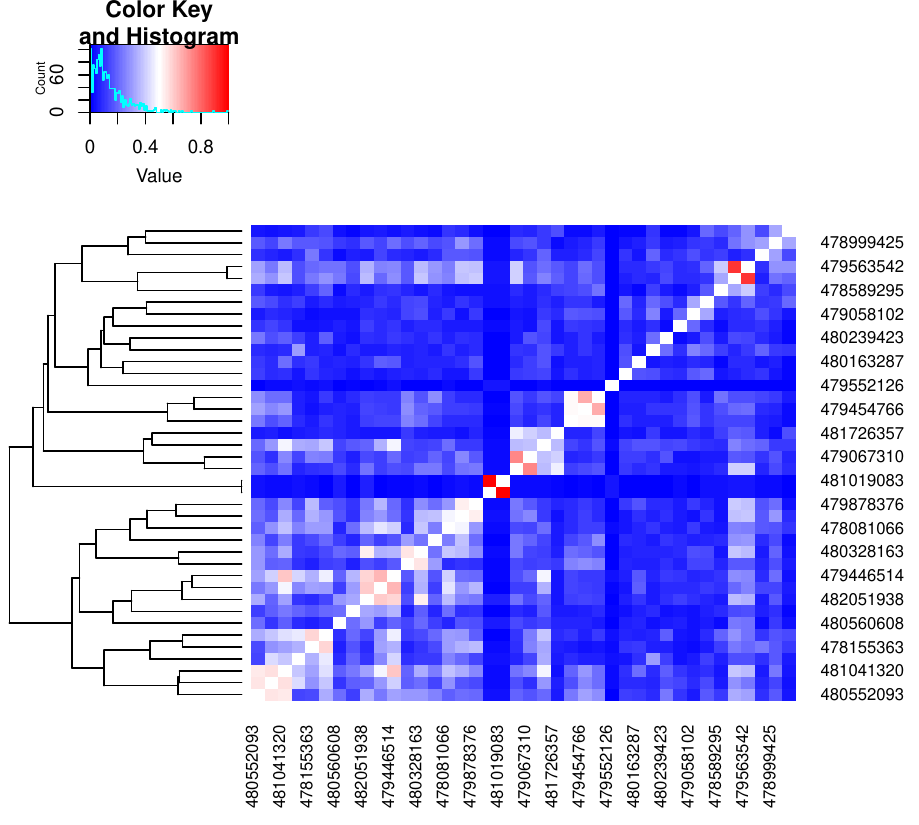}
\caption{Q4 similarity}
\label{fig:q4_similarity}
\end{minipage}\hfill
\begin{minipage}{0.42\textwidth}
\centering
\includegraphics[width=\textwidth]{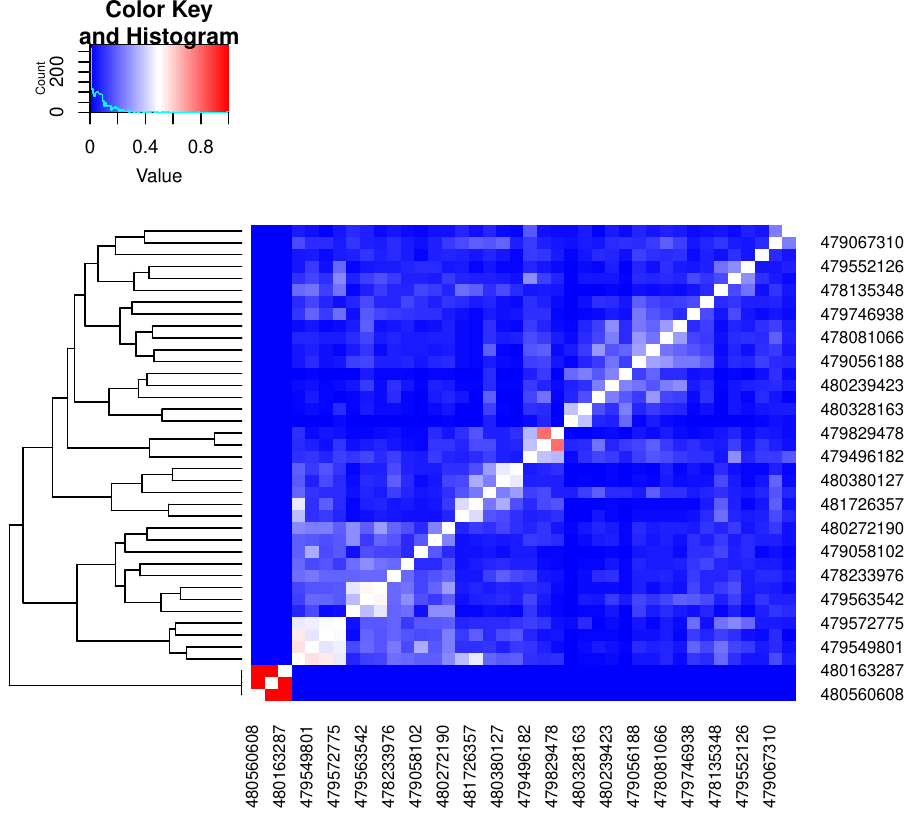}
\caption{Q9 similarity}
\label{fig:q9_similarity}
\end{minipage}
\end{figure}

Previous studies (\cite{Dellacqua:etal:2023}) have found that the use of Gen AI reduces diversity of responses, compared to the responses humans produce without the aid of Gen AI. To quantitatively assess the similarity of answers among trial participants using Gen AI, we generated heatmaps from a similarity matrix alongside a hierarchical clustering tree diagram, which allowed for a visual representation of answer consistency and variation. Figure \ref{fig:q4_similarity} and figure \ref{fig:q9_similarity} show the heatmaps for questions 4 and 9. In question 4 where the treatment group using Gen AI obtained higher quality scores, we see more similar answers in the group. Conversely, for question 9 where the treatment group scored more poorly, there is less similarity in the answers provided by the group. This could be due to a combination of reasons. Firstly, the Gen AI tool may have been unable to collate the information required to answer the question from throughout the document. Secondly, the feedback gathered during the post-trial survey indicated that participants often modified the questions, used multiple prompts, or modified the responses provided by the Gen AI tool to formulate their final answer. This diverse approach in interacting with Gen AI likely contributed to the variation in responses.

The findings from this trial highlight both the strengths and limitations of Gen AI in document comprehension tasks. The treatment group's strong performance on questions with easily identifiable information showcases the potential of Gen AI to streamline certain types of queries, offering valuable support in retrieving targeted information efficiently. However, the variability in performance on more complex questions underscores the current limitations when collating scattered information. The variability in responses reinforces the need for human oversight. While Gen AI can act as a powerful aid, assisting with quicker access to information and offering initial insights, it is not a silver bullet solution. Users must remain engaged in the process, refining queries, verifying outputs and ensuring completeness. 

\subsection{Data}

In this section we provide some observations on the performance of the participants on individual questions in the Data task with a particular focus on the treatment group and their use of the Gen AI application.

\begin{figure}[htbp]
\centering
\begin{minipage}{0.32\textwidth}
\centering
\includegraphics[width=\textwidth]{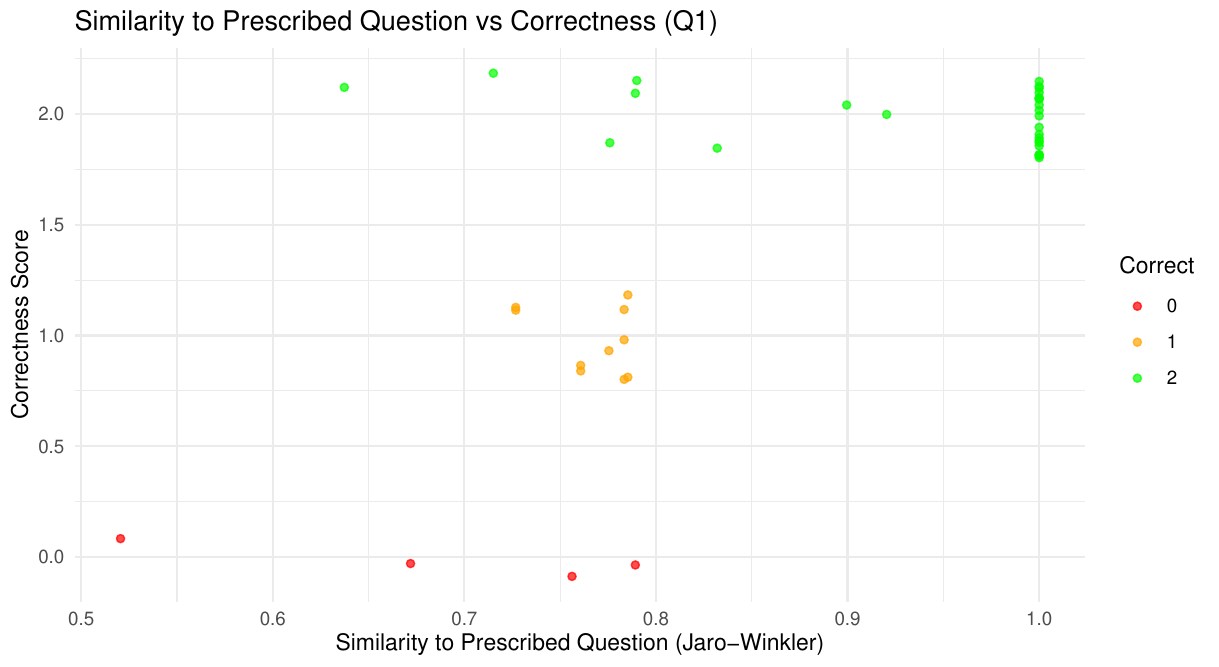}
\caption{Q1 Data experiment task}
\label{fig:Q1_data}
\end{minipage}\hfill
\begin{minipage}{0.32\textwidth}
\centering
\includegraphics[width=\textwidth]{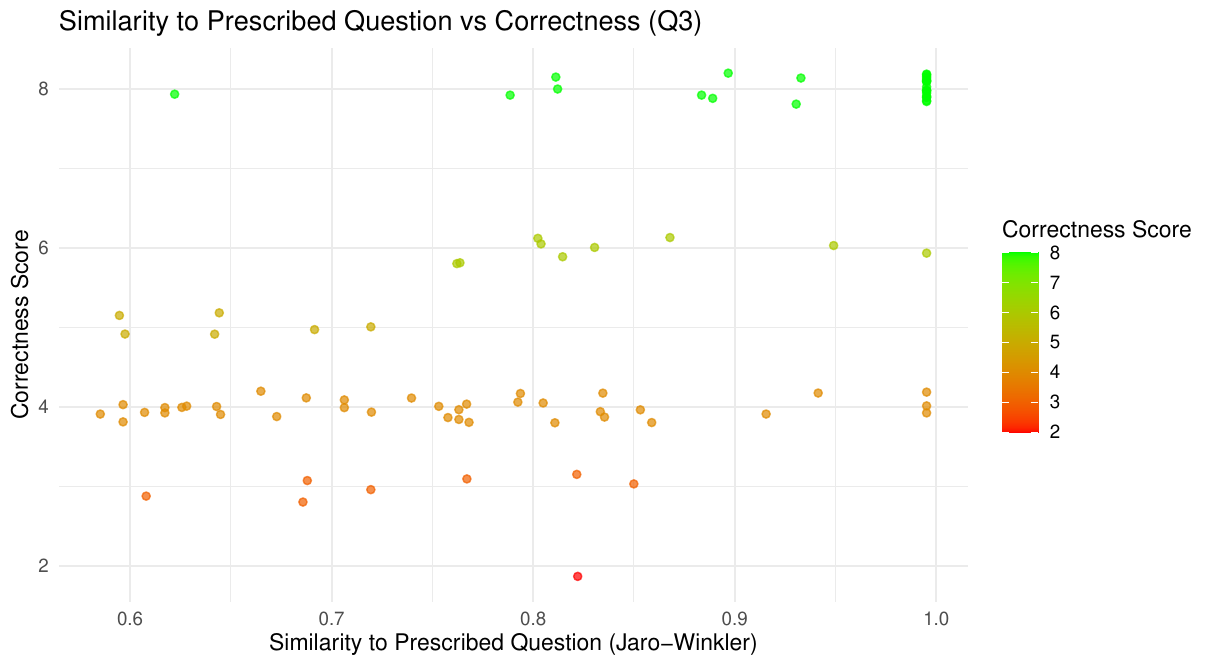}
\caption{Q3 Data experiment task}
\label{fig:Q3_data}
\end{minipage}\hfill
\begin{minipage}{0.32\textwidth}
\centering
\includegraphics[width=\textwidth]{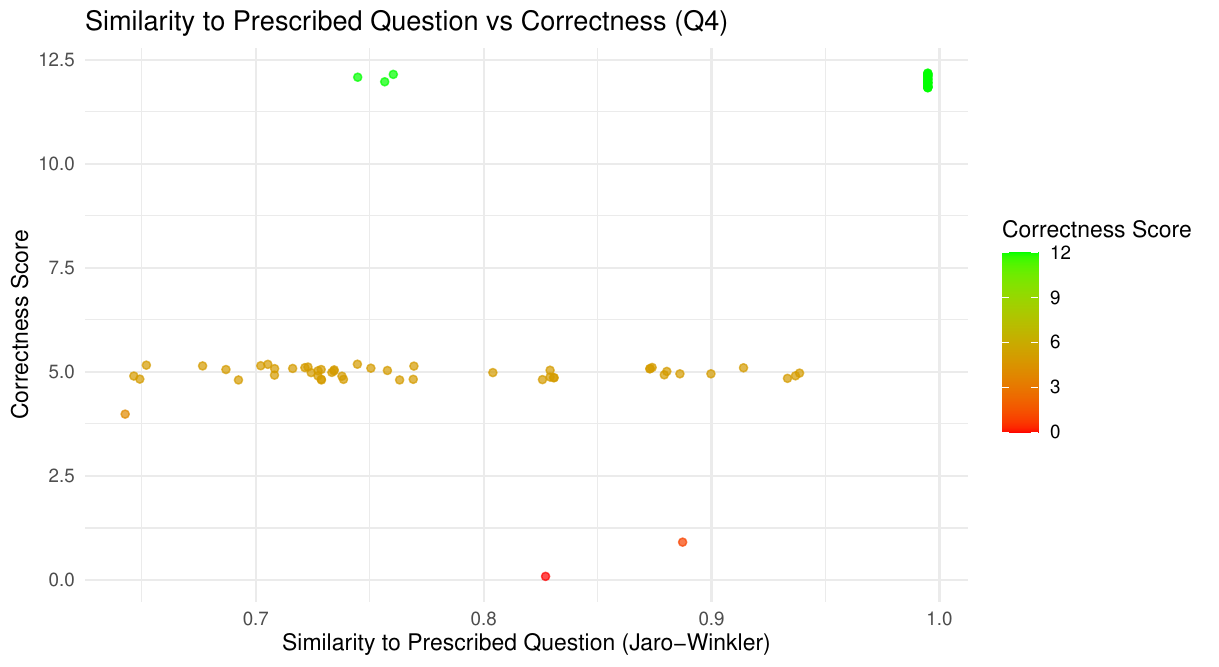}
\caption{Q4 Data experiment task}
\label{fig:Q4_data}
\end{minipage}
\end{figure}

In two of the questions, we observed that some participants omitted key words in the question submitted to the application, resulting in an incorrect output being generated. An example of this is in question 1 (“Has the Solvency Capital Requirement (SCR) ratio for the firm changed from 2021 to 2022”) where the word ‘ratio’ was omitted from the query. Another example of this is in question 3 (“What were the Tiers of Total eligible Own Funds to meet the SCR from 2021 to 2022 for the firm and how have they changed?”) where the exclusion of the word ‘SCR’ resulted in both MCR and SCR being considered and a larger than expected answer returned. Figures \ref{fig:Q1_data} and \ref{fig:Q3_data} demonstrate that prompts with greater similarity to the prescribed question resulted in higher quality scores compared to prompts that differed from the prescribed question.

A final observation is that participants submitted questions in different ways to the Gen AI application, affecting the secondary outcome variable (task completion time). For example in question 4 (“What is the breakdown of direct business Gross Written Premium (GWP) by SII line of business in 2022 for the firm?”). The GWP for twelve lines of business were required in the answer for this question. Submitting the original question to the application produced the correct result in a single response. However, some participants submitted questions to obtain the GWP for individual lines of business, which resulted in time lost on this question. Figure \ref{fig:Q4_data} demonstrates that many participants differed from the prescribed question, and quality scores were generally lower for these participants. This suggests that participants weren’t aware of the possibility of asking the question in a more general sense to receive all of the information required for the answer, and that further prompt training would help improve this.

\clearpage
\section{Appendix - Regression Tables}

This section contains the regression tables, examining the impact of using Gen AI on the quality outcome variable across the three model specifications. Column 1 uses only the treatment variable, while Columns 2 and 3 progressively incorporate additional covariates, including enrolment data and the baseline assessment. 

\subsection{Documents}
\label{sec:documentsregression}

\begin{table}[H]
\centering    
\begin{minipage}{0.85\textwidth}
\includegraphics[width=\textwidth]{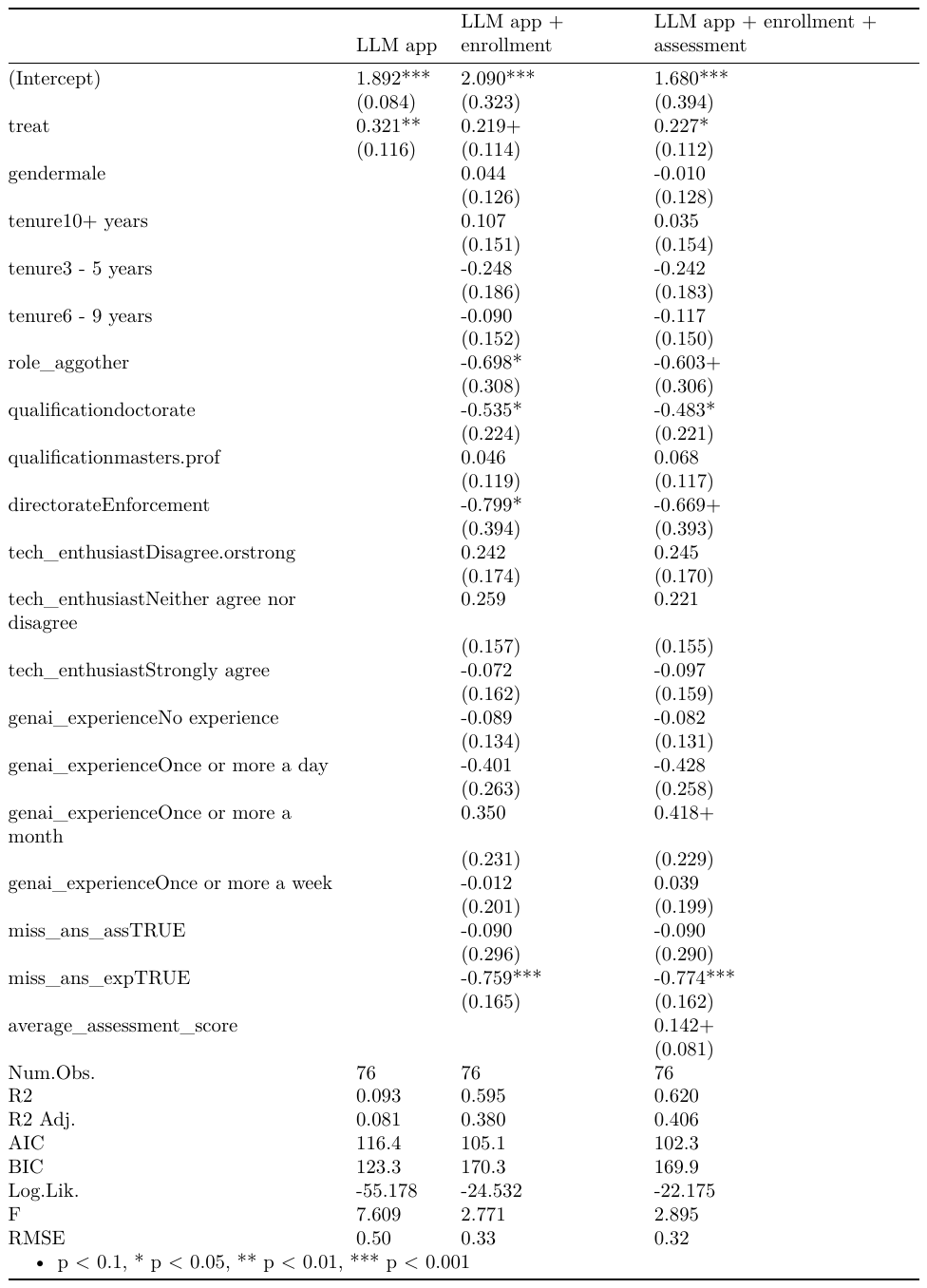}
\caption{Documents Model Summary - Quality}
\label{docsregression}
\end{minipage}
\end{table}\hfill 

\newpage

\begin{table}[H]
\centering    
\begin{minipage}{0.85\textwidth}
\includegraphics[width=\textwidth]{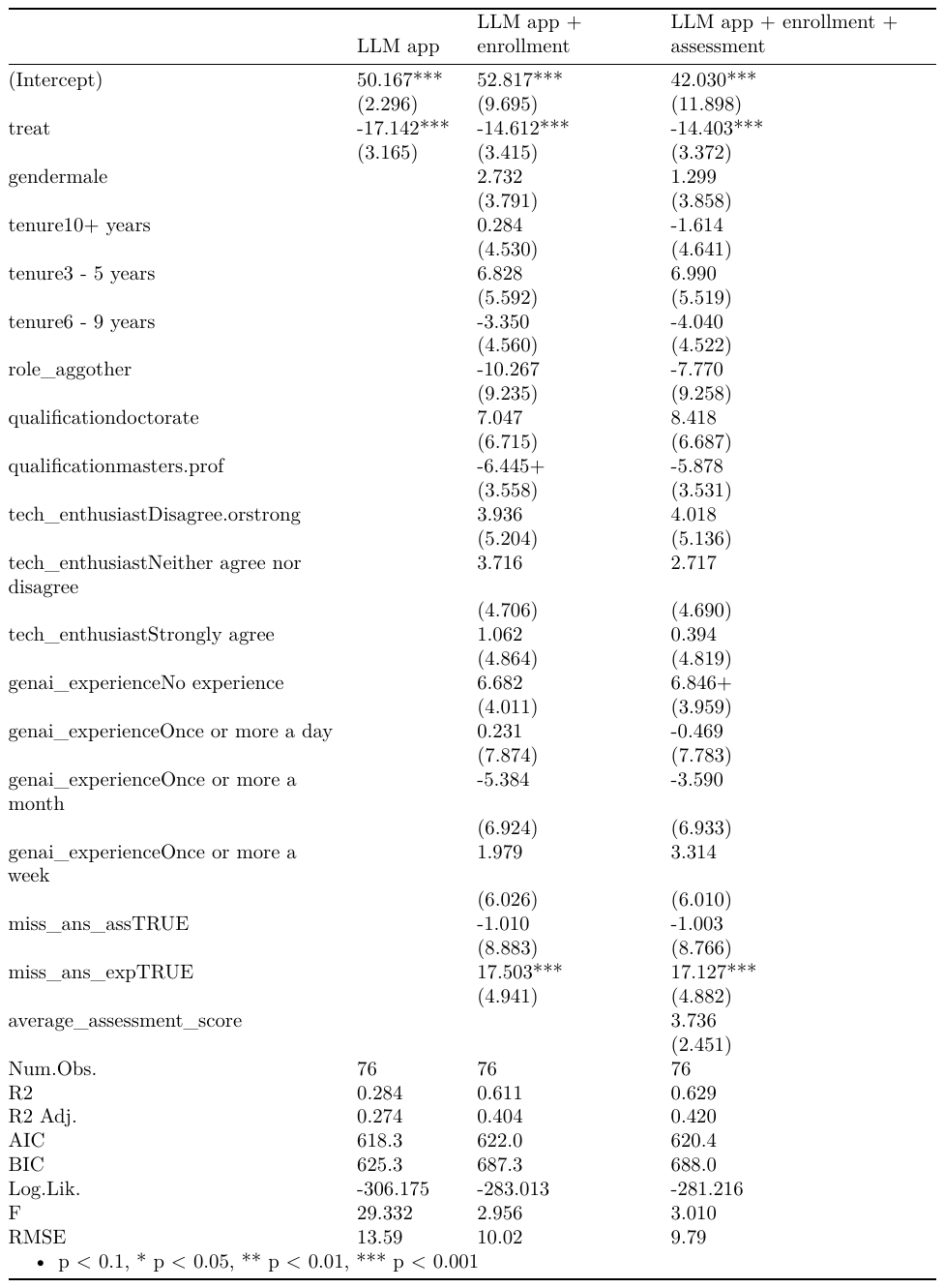}
\caption{Documents Model Summary - Time}
\label{docsregressiontime}
\end{minipage}
\end{table}\hfill 

\newpage

\subsection{Data}
\label{sec:dataregression}

\begin{table}[H]
\centering    
\begin{minipage}{0.85\textwidth}
\includegraphics[width=\textwidth]{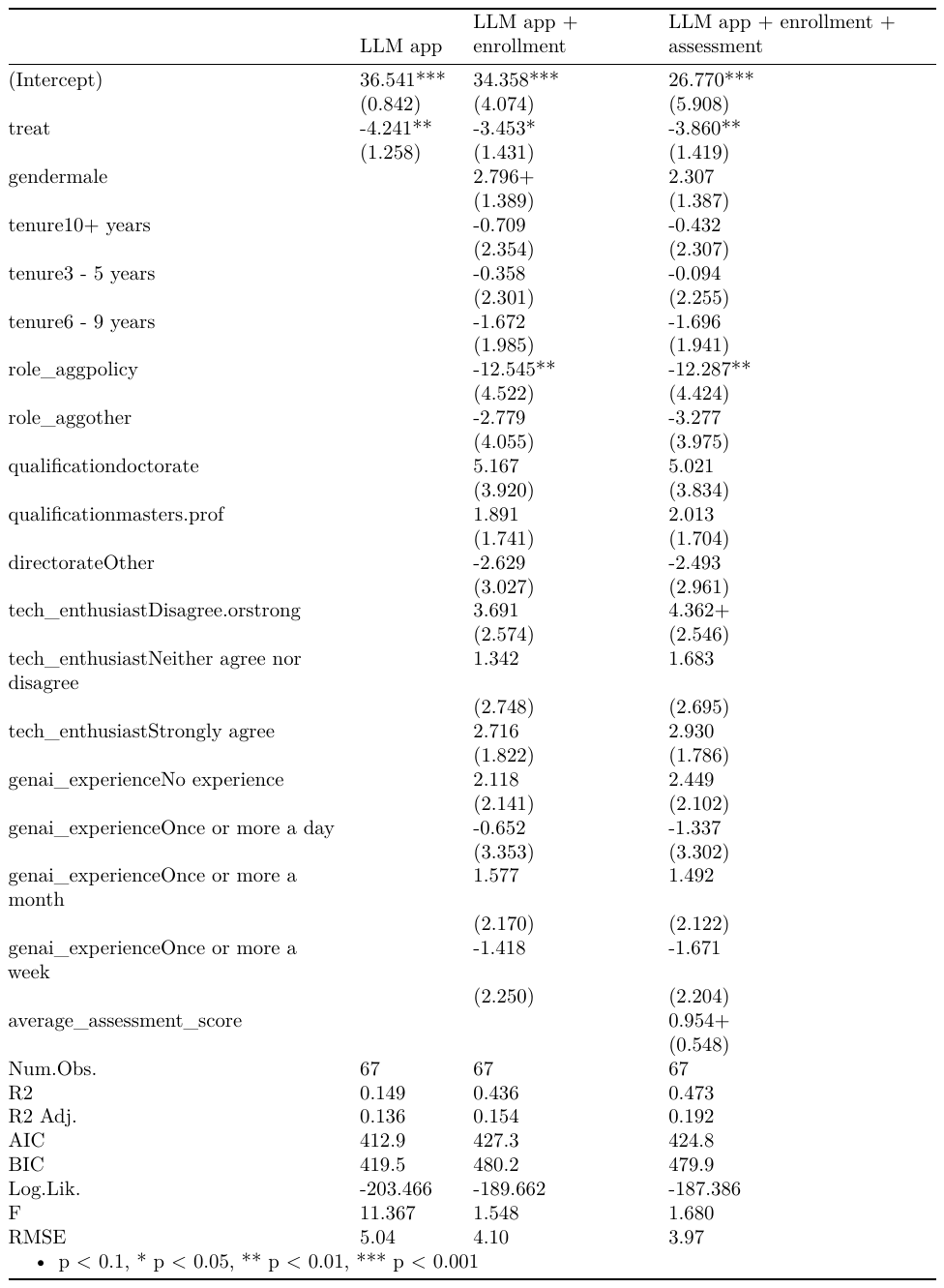}
\caption{Data Model Summary - Quality}
\label{dataregression}
\end{minipage}
\end{table}\hfill 

\newpage

\begin{table}[H]
\centering    
\begin{minipage}{0.85\textwidth}
\includegraphics[width=\textwidth]{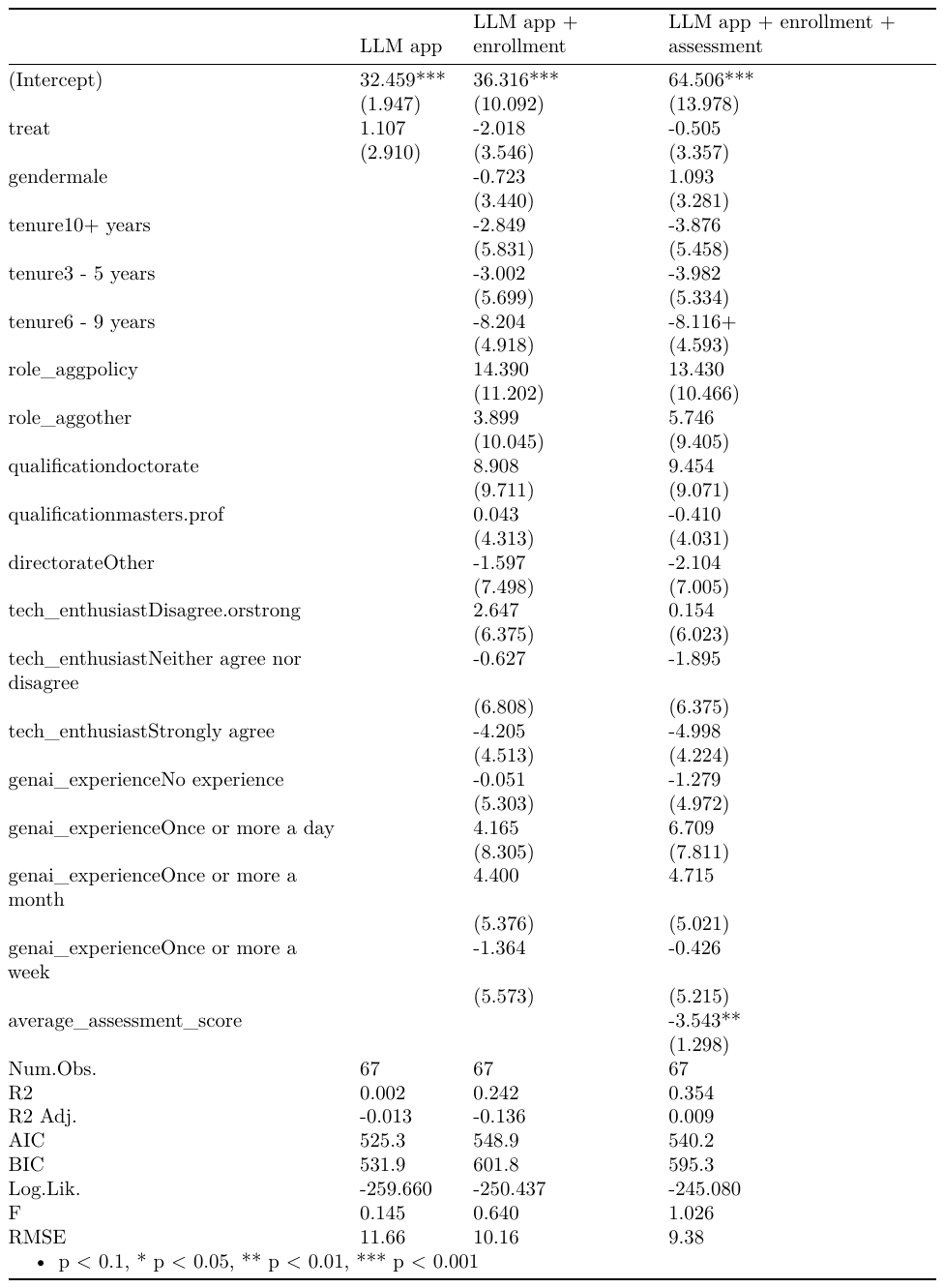}
\caption{Data Model Summary - Time}
\label{dataregressiontime}
\end{minipage}
\end{table}\hfill 

\newpage
\section{Appendix - Feedback Survey}
\label{sec:feedbacksurvey}

To gain insights into participant experiences, a feedback survey was conducted following the trials. There was a 29\% response rate. Charts summarising the main results regarding participants' use of the Gen AI applications follow. 

\begin{figure}[htbp]
\centering
\begin{minipage}{0.45\textwidth}
\centering
\includegraphics[width=\textwidth]{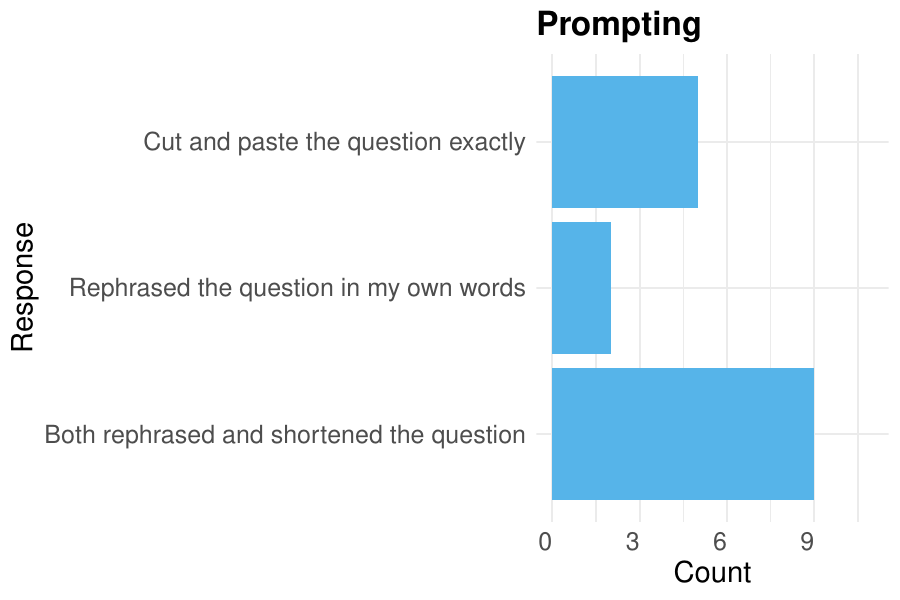}
\caption{Documents Prompting}
\label{fig:docs_feedback_prompt}
\end{minipage}\hfill
\begin{minipage}{0.45\textwidth}
\centering
\includegraphics[width=\textwidth]{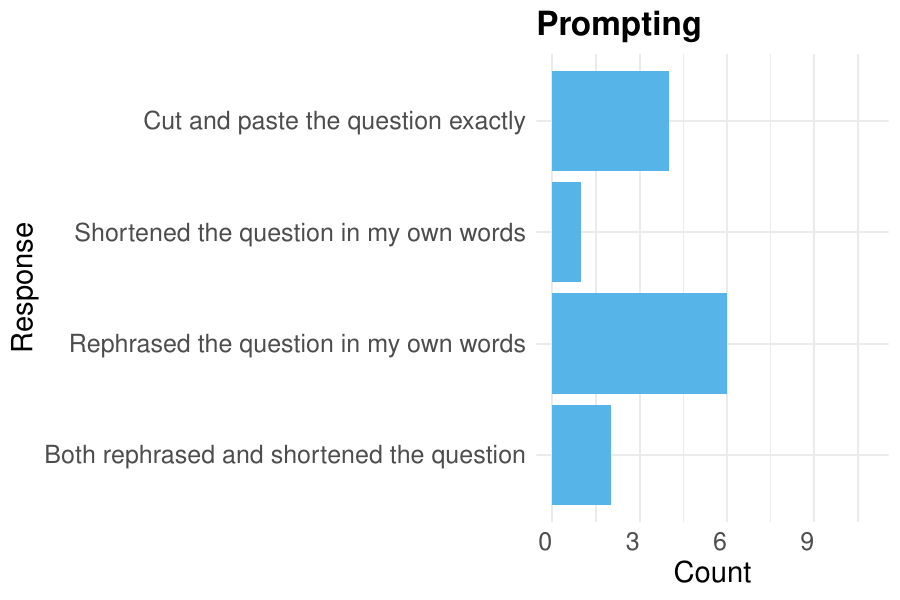}
\caption{Data Prompting}
\label{fig:data_feedback_prompt}
\end{minipage}
\end{figure}

\begin{figure}[htbp]
\centering
\begin{minipage}{0.45\textwidth}
\centering
\includegraphics[width=\textwidth]{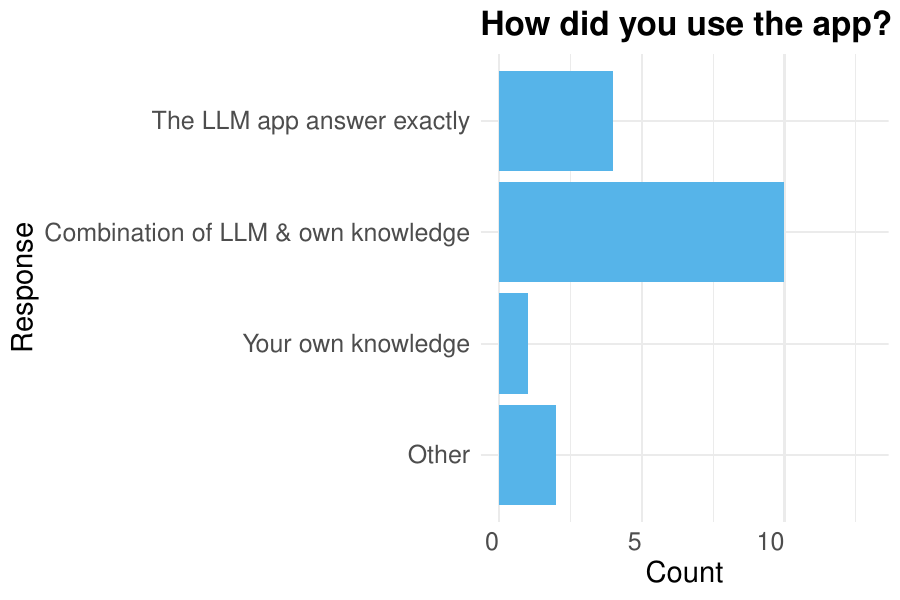}
\caption{Documents Answering}
\label{fig:docs_feedback_answer}
\end{minipage}\hfill
\begin{minipage}{0.45\textwidth}
\centering
\includegraphics[width=\textwidth]{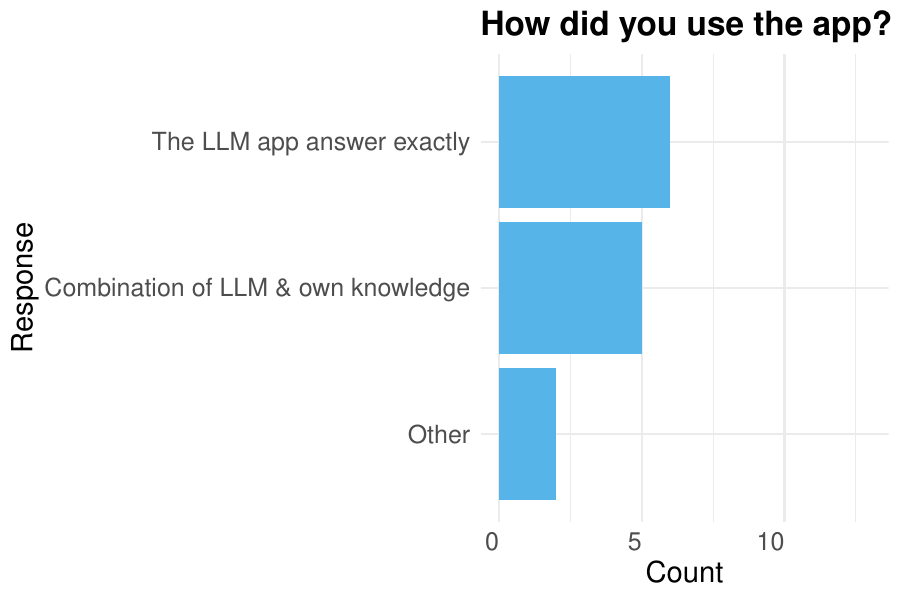}
\caption{Data Answering}
\label{fig:data_feedback_answer}
\end{minipage}
\end{figure}

\newpage
\section{Appendix - Applications}
\label{sec:apps}

This section outlines the technical design of the applications used in the experiment.

\subsection{Documents task}

The participants in the treatment group of the Documents task were provided with a chat-bot web application based on a Retrieval Augmented Generation (RAG) pipeline. The documents used for the tasks were chunked and converted into embeddings using default settings in Azure AI studio (at the time of the experiment - fixed-size chunking and Open AI’s \textit{text-embedding-ada-002} embedding model). The application then took the user queries as inputs and used hybrid (vector + keyword) search to determine the document chunks most relevant to the user's query (Retrieval). These chunks were passed to a deployment of the GPT-4o Large Language Model (Augmentation). Finally, the Large Language Model generated output based on the retrieved document chunks (Generation) and returned this to the user as the response in the web application.

\subsection{Data task}

The participants in the treatment group of the Data task were provided with a Natural-Language-to-SQL web application.
The web application was implemented based on the Vanna project (https://vanna.ai/). A SQLite database was set up local to the web application containing 4 years of publicly available SFCR data. The GPT-4o Large Language Model was prompted with the Data Definition Language of this SQLite database, and it was also prompted with the Solvency II taxonomy to help understand the structure of the database. Twenty example Question-SQL response pairs were provided to the web application (Table \ref{tab:questions_sql_answers} provides examples of these pairs). The Vanna Natural-Language-to-SQL engine is set up in such a way that only the most relevant taxonomy information and sample Question-SQL pairs are transmitted to the model to economize on input token usage.

\begin{table}[ht]
    \centering
    \begin{tabular}{|p{4cm}|p{10cm}|}
        \hline
        \textbf{Question} & \textbf{SQL-Answer} \\
        \hline
        For the firm, how much of their Liabilities is composed of Technical provisions over time. &
        \scriptsize{\texttt{SELECT year\_end, InstitutionName, SUM(CASE WHEN row\_code IN ('R0510', 'R0600', 'R0690') THEN CAST(Euro\_value\_or\_percentage\_value AS FLOAT) ELSE 0 END) AS TechnicalProvisions, (SELECT CAST(Euro\_value\_or\_percentage\_value AS FLOAT) FROM Insurance\_data WHERE InstitutionName = id.InstitutionName AND year\_end = id.year\_end AND row\_code = 'R0900' AND templates LIKE 'S.02.01\%') AS TotalLiabilities, (SUM(CASE WHEN row\_code IN ('R0510', 'R0600', 'R0690') THEN CAST(Euro\_value\_or\_percentage\_value AS FLOAT) ELSE 0 END) / (SELECT CAST(Euro\_value\_or\_percentage\_value AS FLOAT) FROM Insurance\_data WHERE InstitutionName = id.InstitutionName AND year\_end = id.year\_end AND row\_code = 'R0900' AND templates LIKE 'S.02.01\%')) * 100 AS PercentageOfLiabilities FROM Insurance\_data id WHERE InstitutionName LIKE '\%INSURER NAME\%' AND templates LIKE 'S.02.01\%' AND row\_code IN ('R0510', 'R0600', 'R0690', 'R0900') GROUP BY year\_end, InstitutionName ORDER BY year\_end;}} \\
        \hline
        What is the total value of different investment types for the firm over the years, and how do they rank annually? &
        \scriptsize{\texttt{SELECT year\_end, row\_name AS InvestmentType, SUM(CAST(Euro\_value\_or\_percentage\_value AS FLOAT)) AS InvestmentValue FROM Insurance\_data WHERE InstitutionName LIKE '\%INSURER\_NAME\%' AND templates LIKE 'S.02.01\%' AND row\_code IN ('R0080', 'R0090', 'R0110', 'R0120', 'R0140', 'R0150', 'R0160', 'R0170', 'R0180', 'R0190', 'R0200', 'R0210', 'R0220') GROUP BY year\_end, row\_name ORDER BY year\_end, InvestmentValue DESC;}} \\
        \hline
        Who had the most property in 2022? &
        \scriptsize{\texttt{SELECT InstitutionName, SUM(CAST(Euro\_value AS FLOAT)) AS TotalProperty FROM Insurance\_data WHERE year\_end = 2022 AND row\_code IN ('R0040', 'R0050') -- Investment property and Property held for own use AND templates LIKE 'S.02.01\%' GROUP BY InstitutionName ORDER BY TotalProperty DESC LIMIT 1;}} \\
        \hline
        What institutions wrote the most premium in France in 2021? &
        \scriptsize{\texttt{SELECT InstitutionName, SUM(CAST(Euro\_value AS FLOAT)) AS TotalPremiums FROM Insurance\_data WHERE col\_code = 'FR' AND year\_end = 2021 AND (row\_code = 'R0110' OR row\_code = 'R0120' OR row\_code = 'R0130' OR row\_code = 'R1410') AND templates LIKE 'S.05.02\%' GROUP BY InstitutionName ORDER BY TotalPremiums DESC LIMIT 1;}} \\
        \hline
        What were the firm's three largest lines of business by average annual GWP from 2020 to 2022? &
        \scriptsize{\texttt{SELECT col\_name AS LineOfBusiness, AVG(CAST(Euro\_value\_or\_percentage\_value AS FLOAT)) AS AverageGWP FROM Insurance\_data WHERE InstitutionName LIKE '\%INSURER\_NAME\%' AND year\_end BETWEEN 2020 AND 2022 AND (row\_code = 'R0110' OR row\_code = 'R0120' OR row\_code = 'R0130' OR row\_code = 'R1410') AND templates LIKE 'S.05.01\%' GROUP BY col\_name ORDER BY AverageGWP DESC LIMIT 3;}} \\
        \hline
    \end{tabular}
    \caption{Question and SQL-Answer Pairs}
    \label{tab:questions_sql_answers}
\end{table}

\bibliographystyle{unsrt}  
\newpage
\bibliography{references}

\end{document}